\documentclass[10pt, twocolumn,secnumarabic,amssymb, nobibnotes, superscriptaddress, nofootinbib,aps,prd]{revtex4-2}
\usepackage[colorlinks,citecolor=blue,linkcolor=blue,anchorcolor=blue,filecolor=blue, urlcolor=blue]{hyperref}
\usepackage{natbib}

\usepackage{amsmath}
\usepackage{adjustbox}
\usepackage{graphicx}
\usepackage{booktabs}       
\usepackage{amsfonts}       
\usepackage{nicefrac}       
\usepackage{microtype}
\usepackage{cleveref}
\usepackage{xcolor}
\usepackage{multirow}
\usepackage{makecell}
\usepackage{array}
\usepackage{orcidlink}
\usepackage{float}

\begin{document}

\author{Adamu Issifu \orcidlink{0000-0002-2843-835X}} 
\email{ai@academico.ufpb.br}
\affiliation{Departamento de F\'isica, Instituto Tecnol\'ogico de Aeron\'autica, DCTA, 12228-900, S\~ao Jos\'e dos Campos, SP, Brazil} 
\affiliation{Laborat\'orio de Computa\c c\~ao Cient\'ifica Avan\c cada e Modelamento (Lab-CCAM), Brazil}
\affiliation{CFisUC, Department of Physics, University of Coimbra, 3004-516 Coimbra, Portugal}

\author{Andreas Konstantinou~\orcidlink{0000-0002-1072-7313}}\email{akonst29@ucy.ac.cy}
\affiliation{Department of Physics, University of Cyprus, P.O. Box 20537, 1678 Nicosia, Cyprus}
\affiliation{Computation-based Science and Technology Research Center,
The Cyprus Institute, 20 Kavafi Str., Nicosia 2121, Cyprus}

\author{Prashant Thakur~\orcidlink{0000-0001-5930-7179}}
\email{prashant@yonsei.ac.kr}
\affiliation{Department of Physics, Yonsei University, Seoul, 03722, South Korea}

\author{Tobias Frederico \orcidlink{0000-0002-5497-5490}} 
\email{tobias@ita.br}

\affiliation{Departamento de F\'isica, Instituto Tecnol\'ogico de Aeron\'autica, DCTA, 12228-900, S\~ao Jos\'e dos Campos, SP, Brazil} 
\affiliation{Laborat\'orio de Computa\c c\~ao Cient\'ifica Avan\c cada e Modelamento (Lab-CCAM), Brazil}

\title{Rotational enhancement and stability of protoquark stars during thermal evolution}

\begin{abstract}
 We present the first systematic study of rigidly rotating protoquark stars based on isentropic equations of state (EOS) within the density-dependent quark mass (DDQM) framework. Using a quasi-static equilibrium approach, we follow the Kelvin--Helmholtz evolution from hot, lepton-rich matter to a cold, catalyzed quark star (QS). Rotation substantially enhances the maximum stable mass (by up to $\sim 40\%$), equatorial radius, and key rotational observables, with the ratio of rotational kinetic to gravitational potential energy, $T_{\rm kin}/|W|$, reaching $0.18$--$0.19$ near the Keplerian limit, indicating a heightened susceptibility to gravitational-wave--emitting instabilities. Thermal evolution introduces a clear ordering: all stellar properties peak during the lepton-rich stages and decrease monotonically as the star cools. Compared to hadronic stars, rotating proto-QSs exhibit larger radii, higher moments of inertia, and stronger quadrupolar deformation, producing a distinct signature in the mass--radius--spin plane. The EOS parameters are constrained using current astrophysical observations, including mass–radius measurements from HESS~J1731--347 and PSR~J0030+0451, the high-mass constraint from PSR~J0740+6620, and mass–radius constraints inferred from GW170817. The results demonstrate that future multimessenger observations must account for both thermal history and rotation to identify quark matter (QM) in compact stars robustly.

\end{abstract}

\maketitle

\section{Introduction}
The theoretical possibility of compact objects composed of self-bound up ($u$), down ($d$), and strange ($s$) quarks has attracted long-standing interest in nuclear astrophysics \cite{Witten:1984rs, 1986ApJ...310..261A}. The existence of such objects is motivated by the conjecture that strange QM may constitute the true ground state of nuclear matter \cite{Bodmer:1971we, PhysRevD.30.2379, Weber:2004kj}. On this basis, several studies have speculated that some observed pulsars, including Her~X--1, the X-ray burster 4U~1820--30, HESS~J1731--347, and XTE~J1810--197, could be interpreted as QSs \cite{Drake:2002bj, 2022NatAs...6.1444D, Li:1999mk, DiClemente:2022wqp}. A definitive identification of compact objects, however, requires establishing a consistent connection between their internal composition and macroscopic structure through confrontation with observational data. In particular, rotational observables, such as the spin frequency, angular momentum, quadrupole moment, and moment of inertia, provide key diagnostics for distinguishing QSs from hadronic configurations. Early studies of rotating QSs can be found in Refs.~\cite{LattimerApJ1990, StergioulasA&A1999, ZdunikA&A2000, Romanowsky:2000zb, Gondek-Rosinska:2000rjg}.

Early studies of rotating strange stars, based on simplified EOS that neglected quark masses and interactions, predicted Keplerian frequencies comparable to or even lower than those of neutron stars (NSs)~\cite{LattimerApJ1990, ZdunikPhRvD1990, PrakashPhLB1990e}, reflecting their relatively low maximum masses. In contrast, modern EOS incorporating quark masses and interactions, including those adopted here, support substantially higher maximum masses and correspondingly larger rotational frequencies~\cite{Gondek-RosinskaA&A2000, BhattacharyyaMNRAS2016}. Consequently, the rotational properties of self-bound QSs differ qualitatively from those of NSs. Uniform rotation can increase the maximum mass of a QS by more than $40\%$~\cite{Gourgoulhon:1999vx, Gondek-RosinskaA&A2000, ZhouAN2017}, compared with a typical $\sim20\%$ enhancement for NSs, while differential rotation can sustain even more massive configurations~\cite{SzkudlarekASPC2012, Szkudlarek:2019odl, Zhou:2019hyy}. Moreover, the ratio of rotational kinetic to gravitational binding energy, $T_{\rm kin}/|W|$, is significantly larger for QSs~\cite{Gourgoulhon:1999vx, Gondek-RosinskaA&A2000}, indicating an enhanced susceptibility to nonaxisymmetric instabilities, although triaxial deformation may ultimately limit their attainable spin rates~\cite{ZhouJPhCS2017, ZhouPhRvD2018}.
                                                                                                
In this work, we study the rotational properties of proto-QSs within a quasi-static approximation, in which the thermodynamic conditions, such as the entropy per baryon $s_B$ and the lepton fraction $Y_l$, are fixed at each evolutionary stage. Rotating equilibrium configurations are constructed using the open-source \texttt{rns} code \cite{Stergioulas95}. The underlying EOS is modeled within the DDQM framework, with parameters selected from Table~II of \cite{daSilva:2023okq} in order to maximize the stellar mass. The thermodynamics of hot QM follows the formalism introduced in Sec.~IIB of \cite{Issifu:2023qoo}, which focuses on static proto-QSs. The evolution is represented by a sequence of equilibrium snapshots corresponding to four distinct stages. Within this framework, we compute the mass--radius relation, the distribution of the core temperature along stellar sequences, and several key rotational observables. These include the mass as a function of spin frequency, the moment of inertia, angular momentum, polar redshift, and quadrupole moment. A central result of this study is the complete energy decomposition of rotating proto-QSs, which provides complementary information of direct relevance for observational missions~\cite{Issifu:2025hqg}.

The present study is astrophysically motivated by the quark--nova scenario proposed by \cite{Ouyed_2002}, which provides a possible formation channel for hot, rapidly rotating proto--QSs. In this picture, an NS undergoes an explosive transition to deconfined QM, producing a neutrino-trapped, high-entropy, and rapidly rotating quark star immediately after the phase conversion. The subsequent Kelvin--Helmholtz evolution proceeds approximately through quasi-isentropic cooling and deleptonization stages, naturally connecting the hot lepton-rich configurations studied here to the final cold catalyzed QS configurations that may be probed observationally. Moreover, possible links between quark novae and high-energy transients, such as gamma-ray bursts \cite{Ouyed:2001ts,Ouyed:2001cg} and superluminous supernova-like events \cite{Ouyed_2020}, further motivate the study of rapidly rotating proto--QSs, since their thermal and rotational properties may leave observable imprints on the associated electromagnetic, neutrino, and gravitational-wave signals \cite{universe8060322}. In this context, our work provides a framework connecting the DDQM EOS to the early evolution of newly born QSs.

We adopt two EOS parameterizations from \cite{daSilva:2023okq}, constrained by recent astrophysical observations through Bayesian inference. To our knowledge, this work presents the first systematic investigation of rotating proto-QSs based on isentropic EOS. Earlier studies of rotating hot compact QSs employing isothermal EOS can be found in~\cite{Shen:2005vh}. The possible role of strange QM in the early universe has also been explored in \cite{Madsen:1998uh}. A full general relativistic simulation of rapidly rotating QSs, including oscillation modes and universal relations, can also be found in \cite{Chen:2023bxx}.

This paper is organized as follows. In \cref{md}, we introduce the model formalism and the equation of state, which is discussed in detail in \cref{md1} and separated into the cold (\cref{md2}) and hot (\cref{md3}) QM sectors. The formalism used to describe the rotational properties is presented in \cref{rp}. The results and their analysis are given in \cref{rvan}, and our conclusions are summarized in \cref{conc}.

\section{Model Description and Formalism} \label{md}

\subsection{The equation of state}\label{md1}
The presentation of the EOS is divided into two parts to fully investigate the star’s evolutionary process. We first discuss the cold, catalyzed QM configuration, followed by the finite-temperature protoquark star phase. The EOS that forms the foundation of this study is derived from \cite{Issifu:2023qoo}, where the effective temperature was introduced through a phenomenological method. The appropriate coupling constants were also adopted from \cite{daSilva:2023okq}, in which the authors determined the free model parameters of the DDQM using Bayesian inference in light of recent astrophysical data.

\subsubsection{Cold quark matter}\label{md2}
The QM EOS is described using the density–dependent quark mass (DDQM) approach. In this framework, the interaction among quarks is encoded through the baryon density $\rho_{B}$, which determines how the quark masses vary with the medium. The density dependence is introduced via the ansatz~\cite{Xia:2014zaa, Issifu_2025}:
\begin{equation}
    m_i = m_{i0} + \frac{D}{\rho_B^{1/3}} + C\,\rho_B^{1/3},
\end{equation}
where $m_{i0}$ ($i=u,d,s$) denotes the current quark masses and $C$ and $D$ are model parameters. Although widely used, the DDQM model is known to violate thermodynamic consistency. A standard remedy is to replace the free chemical potential $\mu_i$ with an effective one, $\mu_i^{*}$, which explicitly depends on $\rho_B$, and to employ the density-dependent quark masses $m_i(\rho_B)$ in the thermodynamic potential. The free-energy density is then expressed as
\begin{equation}
    f = \Omega_0\!\left(\{\mu_i^{*}\},\{m_i\}\right)
        + \sum_i \mu_i^{*} n_i,
\end{equation}
where the thermodynamic potential takes the form (see ~\cite{Issifu:2023qoo} and references therein)
\begin{equation}
    \Omega_0 = - \sum_i \frac{\gamma_i}{24\pi^2}
    \left[
        \mu_i^{*}\nu_i\!\left(\nu_i^2 -\frac{3}{2}m_i^2\right)
        + \frac{3}{2} m_i^4
          \ln\!\left(\frac{\mu_i^{*}+\nu_i}{m_i}\right)
    \right],
\end{equation}
with $\gamma_i = 6$ accounting for spin and color degeneracy.  
The corresponding Fermi momentum is
\begin{equation}
    \nu_i = \sqrt{\mu_i^{*2}-m_i^{\,2}},
\end{equation}
leading to the number density
\begin{equation}
    \rho_i
    = \frac{\gamma_i}{6\pi^2}
      \left(\mu_i^{*2}-m_i^{\,2}\right)^{3/2}
    = \frac{\gamma_i\,\nu_i^{3}}{6\pi^2}.
\end{equation}

The physical chemical potential is related to its effective counterpart through
\begin{equation}
    \mu_i = \mu_i^{*} - \mu_I,
\end{equation}
where $\mu_I$ encodes the interaction shift. Under $\beta$-equilibrium, the chemical potentials satisfy
\begin{equation}
    \mu_u^{*} + \mu_e = \mu_d^{*} = \mu_s^{*}.
\end{equation}
The construction of the EOS further requires a global charge neutrality,
\begin{equation}
    \frac{2}{3}\rho_u - \frac{1}{3}\rho_d - \frac{1}{3}\rho_s - \rho_e = 0,
\end{equation}
and baryon number conservation,
\begin{equation}
    \rho_B = \frac{1}{3}(\rho_u+\rho_d+\rho_s).
\end{equation}
The energy density follows from
\begin{equation}
    \varepsilon = \Omega_0 -
    \sum_i \mu_i^{*}
    \frac{\partial \Omega_0}{\partial\mu_i^{*}},
\end{equation}
while the pressure is obtained from
\begin{equation}
    p = -\Omega_0
        + \sum_{i,j}
        \frac{\partial\Omega_0}{\partial m_j}\,
        \rho_i\frac{\partial m_j}{\partial \rho_i}.
\end{equation}

\subsubsection{Hot Quark Matter}\label{md3}
 We begin with the temperature‐dependent free energy density $f$, which serves as the foundation for deriving the energy density $\varepsilon$ and pressure $p$ through a self-consistent thermodynamic framework expressed in terms of the temperature $T$. Hence,
\begin{align}\label{1}
    f &= f(T,V,\{\rho_i\},\{m_i\})\nonumber\\
    &=\Omega_0(T, V,\{\mu^*_i\},\{m_i\}) + \sum_{i=u,d,s}\mu^*_i\rho_i,
\end{align}
where $V$ is the volume of the system. Since the independent state variables do not include the $\mu_i^{*}$ appearing in $\Omega_{0}$, we must establish a link between $\mu_i^{*}$ and the independent variables. This connection is introduced through the particle number density
 $\rho_i$: 
\begin{equation}\label{1a}
    \rho_i = -\dfrac{\partial}{\partial\mu^*_i}\Omega_0(T, V,\{\mu^*_i\},\{m_i\}).
\end{equation}
Taking the derivative of Eq.~(\ref{1}) we have
\begin{equation}\label{2}
    df = d\Omega_0 + \sum_i\rho_id\mu^*_i + \sum_i\mu^*_id\rho_i,
\end{equation}
with 
\begin{equation}\label{3}
    d\Omega_0 = \dfrac{\partial\Omega_0}{\partial T}dT + \sum_i\dfrac{\partial\Omega_0}{\partial\mu^*_i}d\mu^*_i + \sum_i\dfrac{\partial\Omega_0}{\partial m_i}dm_i + \dfrac{\partial\Omega_0}{\partial V}dV,
\end{equation}
and 
\begin{equation}\label{4}
    dm_i = \dfrac{\partial m_i}{\partial T}dT + \sum_j\dfrac{\partial m_i}{\partial \rho_j}d\rho_j.
\end{equation}
Substituting Eqs.~(\ref{3}) and (\ref{4}) into Eq.~(\ref{2}) and grouping the terms, we have
\begin{align}
    df &= \left( \dfrac{\partial\Omega_0}{\partial T} + \sum_i\dfrac{\partial\Omega_0}{\partial m_i}\dfrac{\partial m_i}{\partial T}\right)dT \nonumber\\ &+ \sum_i\left(\mu_i^* +\sum_j\dfrac{\partial\Omega_0}{\partial m_j}\dfrac{\partial m_j}{\partial \rho_i}\right)d\rho_i + \dfrac{\partial \Omega_0}{\partial V}dV.
\end{align}
Comparing the above expression with the thermodynamic relation
\begin{equation}
    df = -SdT + \sum_i\mu_id\rho_i + \left(-P -f + \sum_i\mu_i\rho_i\right)\dfrac{dV}{V},
\end{equation}
we can express the entropy $S$ as:
\begin{equation}\label{4a}
    S = -\dfrac{\partial \Omega_0}{\partial T} - \sum_i \dfrac{\partial m_i}{\partial T}\dfrac{\partial\Omega_0}{\partial m_i},
\end{equation}
the pressure as: 
\begin{equation} \label{5}
    p = -f + \sum_i\mu_i\rho_i - V\dfrac{\partial\Omega_0}{\partial V},
\end{equation}
and the chemical potential in terms of the effective chemical potential is given explicitly as:
\begin{equation} \label{6}
    \mu_i = \mu^*_i + \sum_j\dfrac{\partial \Omega_0}{\partial m_j}\dfrac{\partial m_j}{\partial \rho_i} \equiv \mu_i^* - \mu_I.
\end{equation}
The last term in Eq.~(\ref{5}) persists when the finite size effect of the system cannot be neglected, whether the particle masses are fixed or not. However, it can be ignored in this study just like other previous studies of DDQM \cite{Wen:2005uf, Peng:2000ff, Xia:2014zaa, Backes:2020fyw}, considering an infinitely large system of QM. In this case, the free energy density is independent of $V$, and substituting Eqs.~(\ref{1}) and (\ref{6}) into Eq.~(\ref{5}), the pressure becomes
\begin{align}\label{6a}
    p = -\Omega_0 + \sum_{i,j}\rho_i\dfrac{\partial \Omega_0}{\partial m_j}\dfrac{\partial m_j}{\partial \rho_i}.
\end{align}
The energy density is determined by substituting Eqs.~(\ref{1}) and (\ref{4a}) into $f = \varepsilon - TS$, yielding
\begin{equation}\label{6b}
    \varepsilon = \Omega_0 + \sum_i\mu_i^*\rho_i - T\dfrac{\partial \Omega_0}{\partial T} - T\sum_i \dfrac{\partial m_i}{\partial T}\dfrac{\partial\Omega_0}{\partial m_i}.
\end{equation}
For a given independent variables $T$ and $\rho_i$, the $\mu_i^*$ are found by solving Eq.~(\ref{1a}). All thermodynamic quantities follow from Eqs.~(\ref{4a}), (\ref{6}), (\ref{6a}), and (\ref{6b}). The finite-temperature extension of the quark mass formula is
\begin{multline}\label{m2}
    m_i = m_{i0}+\dfrac{D}{\rho_B^{1/3}} \left(1 + \dfrac{8T}{\Lambda}e^{-\Lambda/T}\right)^{-1}
   \\ + C\rho_B^{1/3}\left(1 + \dfrac{8T}{\Lambda}e^{-\Lambda/T}\right),
\end{multline}
with $\Lambda = 280$ MeV \cite{Deur:2016tte}. For the analysis that follows, we adopt two different EOS parameterizations from \cite{daSilva:2023okq}, where these parameters were determined through Bayesian inference using astrophysical data as constraints. We select \(C = 0.8\) and \(\sqrt{D} = 127.4\,\text{MeV}\), which yields the highest maximum mass in their analysis, and \(C = 0.65\) with \(\sqrt{D} = 133.2\,\text{MeV}\), which produces a maximum mass slightly above the \(2\,M_\odot\) threshold. These parameters are deliberately chosen to yield the maximum possible stable QS mass within the optimized parameter space identified in \cite{daSilva:2023okq}. As shown there, decreasing $C$ or increasing $\sqrt{D}$ relative to the values adopted here results in a smaller maximum stable QSs below the 2\,$\rm M_\odot$ thereshold. The current quark masses follow PDG values: $m_u=2.16$ MeV, $m_d=4.67$ MeV, $m_s=93.4$ MeV \cite{ParticleDataGroup:2022pth}. This temperature-dependent quark mass relation has been adopted in \cite{Wen:2005uf, Chen:2021fdj, Issifu:2023qoo}. {The qualitative interpretation of the $T$-dependence in Eq.~\eqref{m2} is associated with thermal screening effects in the quark--gluon medium. As the temperature increases, the same thermal factor,
\begin{equation}
    f(T)=1+\dfrac{8T}{\Lambda}e^{-\Lambda/T},
\end{equation}
encoding the excitation of quark--antiquark pairs and gluons oppositely modulates the two $\rho_B$-dependent contributions: it enhances the repulsive term proportional to $\rho_B^{1/3}$, increasing the effective quark mass at high densities and stiffening the DDQM EOS through a larger pressure contribution, while suppressing the confining term proportional to $\rho_B^{-1/3}$, thereby screening the color interaction and driving the system toward asymptotic freedom.
    }

\subsection{Rotational Properties}\label{rp}

In this work, rigidly rotating proto-QSs are modeled using the \texttt{rns} code \cite{Stergioulas95}. The code solves Einstein’s field equations for a stationary and axisymmetric spacetime, described by the line element
\begin{align}  \label{eq:metric}
ds^2 &= -e^{{\gamma} + {\lambda}} dt^2 + e^{2{\alpha}} (dr^2 + r^2 d\theta^2) \nonumber\\
  &+ e^{{\gamma} - {\lambda}} r^2 \sin^2\theta (d\phi - {\omega} dt)^2,
\end{align}
where the metric potentials \({\gamma}, {\lambda}, {\alpha} \), and \({\omega} \) depend on the radial coordinate \( r \) and the polar angle \( \theta \). The global stellar quantities, including the gravitational mass \(M\), baryonic mass \(M_0\), angular momentum \(J\), equatorial circumferential radius \(R_e\), and polar redshift \(Z_p\), are computed following Eqs.~(B1), (B2), (B4), (B6), and (B7) of Ref.~\cite{1994ApJ...424..823C}, respectively. The moment of inertia is defined as \(I = J/\Omega\), where \(\Omega = 2\pi\nu\) is the angular velocity and \(\nu\) the spin frequency. The rotational kinetic energy is given by
\begin{equation}
T_{\mathrm{kin}} = \frac{J\Omega}{2},
\end{equation}
while the binding energy is defined as
\begin{equation}
E_{\mathrm{bind}} = M - M_0 .
\end{equation}

The gravitational potential energy is evaluated as
\begin{equation}
|W| = |M_p + T_{\mathrm{kin}} - M| ,
\end{equation}
where the proper mass \(M_p\) is defined according to Eq.~(65) of Ref.~\cite{komatsu1989MNRASa}. The internal energy is then obtained as
\begin{equation}
U = M_p - M_0 .
\end{equation}
The mass quadrupole moment is determined from the asymptotic expansion of the metric functions at large distances. Within the \texttt{rns} framework, the coefficient \(M_2\) is extracted from the \(r^{-3} P_2(\cos\theta)\) term in the expansion of the \(\rho\)-potential \cite{Laarakkers:1997hb,pappas2012multipolemomentsnumericalspacetimes}. The physical quadrupole moment \(Q\) is then obtained as
\begin{equation}
Q = M_2 - \frac{4}{3}\left(\frac{1}{4} + b\right) M^3 ,
\label{eq:Q-final}
\end{equation}
where \(b\) is a dimensionless coefficient determined from the asymptotic behavior of the metric potentials (see Eq.~(9) of Ref.~\cite{Laarakkers:1997hb}), and \(M_2\) corresponds to the \(l=2\) multipole coefficient of the expansion \cite{pappas2012multipolemomentsnumericalspacetimes}. {We compute the percentage (\%) increase in maximum mass enhancement using the relation:
\begin{equation}
\Delta M_{\max}(\%) =
\frac{M_{\max}(r_p/r_e)-M_{\max}(1.0)}{M_{\max}(1.0)} \times 100 \, .
\end{equation}
}

\section{Results and analysis}\label{rvan}

\begin{table*}[t]
\centering
\caption{Stellar properties at maximum mass for different quasi-static evolutionary stages of rotating proto-QSs parameterized by $r_p/r_e$ and ($C = 0.8$ and $\sqrt{D} = 127.4\,\text{MeV}$). The angular momentum $J$, moment of inertia $I$, quadrupole moment $|Q|$, radius $R$, gravitational redshift $Z_p$, the ratio of the rotational kinetic energy $T_{\rm kin}$ to the gravitational potential energy $|W|$, and the spin frequency $\nu$ are determined at the maximum mass $M_{\rm max}$. Also shown are the stellar radius at $1.4\,M_\odot$, $R_{1.4}$, and the temperature at $2.0\,M_\odot$, $T_{2.0}$ ({this is deliberately chosen to amplify the variation of $T$ with $r_p/r_e$, the variations at $M_{\rm max}$ can be found in \cref{MassxT}}).}
\renewcommand{\arraystretch}{1.1}
\small
\resizebox{\textwidth}{!}{
\begin{tabular}{l c c c c c c c c c c c c}
\hline\hline
Model & $r_p/r_e$ & $M_{\max}$ & $R$ & $R_{1.4}$ & $J$ & $I$ & $|Q|$ & $T_{2.0}$ & $\Delta M_{\max}$ & $Z_p$ & $T_{\rm kin}/|W|$ & $\nu$  \\
& & [$M_\odot$] & [km] & [km] & [$10^{49}\rm gcm^2s^{-1}$] & [$10^{45}\rm gcm^2$] & [$10^{42}\rm gcm^2$] & [MeV] & [\%] &  &  & [Hz] \\
\hline

\multirow{6}{*}{$s_B=1,\;Y_l=0.4$}
& 1.0 & 2.33 & 14.71 & 14.92 & 0.00 & $\cdots$  & 0.00 & 9.43  & 0.0  & 0.38 & $\cdots$ & $\cdots$ \\
& 0.9 & 2.45 & 15.39 & 15.52 & 1.79 & 4.95 & 202.49 & 8.90 & 5.2  & 0.42 & 0.03 & 577.08 \\
& 0.8 & 2.59 & 16.40 & 16.21 & 2.93 & 5.90 & 474.62 & 8.31 & 11.2 & 0.47 & 0.07 & 789.94 \\
& 0.7 & 2.79 & 17.82 & 17.04 & 4.29 & 7.47 & 899.78 & 7.67 & 19.7 & 0.53 & 0.12 & 913.96 \\
& 0.6 & 3.03 & 19.73 & 18.08 & 6.02 & 9.79 & 1554.55 & 6.98 & 30.0 & 0.59 & 0.15 & 977.73 \\
& 0.5 & 3.27 & 22.33 & 19.37 & 7.93 & 12.63 & 2417.96 & 6.23 & 40.3 & 0.63 & 0.19 & 999.93 \\
\hline

\multirow{6}{*}{$s_B=2,\;Y_l=0.2$}
& 1.0 & 2.28 & 14.44 & 14.72 & 0.00 & $\cdots$  & 0.00 & 20.32 & 0.0  & 0.38 & $\cdots$ & $\cdots$ \\
& 0.9 & 2.39 & 15.17 & 15.32 & 1.71 & 4.67 & 189.73 & 19.07 & 4.8  & 0.41 & 0.03 & 582.23 \\
& 0.8 & 2.54 & 16.09 & 16.02 & 2.81 & 5.55 & 447.74 & 17.71 & 11.4 & 0.46 & 0.07 & 804.74 \\
& 0.7 & 2.72 & 17.25 & 16.67 & 4.07 & 6.77 & 805.63 & 16.27 & 19.3 & 0.52 & 0.11 & 956.54 \\
& 0.6 & 2.96 & 19.09 & 17.55 & 5.70 & 8.83 & 1383.42 & 14.78 & 29.8 & 0.59 & 0.15 & 1027.09 \\
& 0.5 & 3.20 & 21.89 & $\cdots$ & 7.55 & 11.77 & 2246.61 & $\cdots$ & 40.4 & 0.62 & 0.18 & 1021.57 \\
\hline

\multirow{6}{*}{$s_B=2,\;Y_{\nu_e}=0$}
& 1.0 & 2.24 & 14.26 & 14.68 & 0.00 & $\cdots$  & 0.00 & 22.53 & 0.0  & 0.37 & $\cdots$ & $\cdots$ \\
& 0.9 & 2.35 & 14.98 & 15.27 & 1.65 & 4.44 & 182.64 & 21.07 & 4.9  & 0.41 & 0.03 & 590.25 \\
& 0.8 & 2.49 & 15.88 & 15.97 & 2.70 & 5.27 & 426.80 & 20.06 & 11.2 & 0.46 & 0.07 & 814.72 \\
& 0.7 & 2.67 & 17.11 & 16.82 & 3.92 & 6.50 & 776.60 & 18.15 & 19.2 & 0.52 & 0.10 & 959.49 \\
& 0.6 & 2.90 & 18.89 & 17.88 & 5.46 & 8.40 & 1315.96 & 16.60 & 29.5 & 0.58 & 0.14 & 1035.26 \\
& 0.5 & 3.12 & 21.70 & 19.24 & 7.18 & 11.15 & 2123.04 & 14.93 & 39.3 & 0.61 & 0.18 & 1024.37 \\
\hline

\multirow{6}{*}{$T=0$}
& 1.0 & 2.20 & 14.03 & 14.50 & 0.00 & $\cdots$  & 0.00 &$\cdots$ & 0.0  & 0.37 & $\cdots$ & $\cdots$ \\
& 0.9 & 2.30 & 14.70 & 15.10 & 1.58 & 4.17 & 170.56 & $\cdots$ & 4.5  & 0.41 & 0.03 & 602.17 \\
& 0.8 & 2.44 & 15.65 & 15.80 & 2.59 & 5.01 & 405.79 & $\cdots$ & 10.9 & 0.46 & 0.07 & 824.24 \\
& 0.7 & 2.62 & 16.86 & 16.65 & 3.77 & 6.17 & 737.89 & $\cdots$ & 19.1 & 0.51 & 0.10 & 970.70 \\
& 0.6 & 2.84 & 18.57 & 17.71 & 5.24 & 7.92 & 1238.79 & $\cdots$ & 29.1 & 0.58 & 0.14 & 1052.82 \\
& 0.5 & 3.05 & 21.33 & 18.99 & 6.86 & 10.47 & 1984.78 & $\cdots$ & 38.6 & 0.61 & 0.18 & 1043.31 \\
\hline\hline
\end{tabular}}
\label{Mass_tab}
\end{table*}

\begin{table*}[htbp]
\centering
\caption{Maximum-mass properties for different evolutionary stages and rotational parameter $r_p/r_e$. Here, $M_{\max}$ is given in $M_\odot$, $R(M_{\max})$ and $R_{1.4}$ in km, $J$ in units of $10^{49}\,\mathrm{g\,cm^2\,s^{-1}}$, $I$ in units of $10^{45}\,\mathrm{g\,cm^2}$, and $|Q|$ in units of $10^{42}\,\mathrm{g\,cm^3}$ for EOS parameter \(C = 0.65\) with \(\sqrt{D} = 133.2\,\text{MeV}\).}
\begin{adjustbox}{width=\textwidth}
\begin{tabular}{llcccccccccc}
\hline\hline
Model & $r_p/r_e$ & $M_{\max}$ & $R(M_{\max})$ & $R_{1.4}$ & $J$ & $I$ & $|Q|$ & $\Delta M$ & $Z_p$ & $T_{\mathrm{kin}}/|W|$ & $\nu$ \\
&  & [$M_\odot$] & [km] & [km] & [$10^{49}$] & [$10^{45}$] & [$10^{42}$] & [\%] &  &  & [Hz] \\
\hline

\multirow{6}{*}{$s_B=1,\;Y_l=0.4$}
& 1.0 & 2.1818 & 13.6103 & 14.0139 & 0.00 & \ldots & 0.00 & 0.0 & 0.38 & \ldots & \ldots \\
& 0.9 & 2.2846 & 14.3010 & 14.5741 & 1.56 & 3.98 & 161.28 & 4.7 & 0.41 & 0.03 & 623.06 \\
& 0.8 & 2.4214 & 15.1692 & 15.2488 & 2.56 & 4.74 & 380.83 & 11.0 & 0.46 & 0.07 & 859.62 \\
& 0.7 & 2.5987 & 16.3938 & 16.0261 & 3.72 & 5.90 & 704.53 & 19.1 & 0.51 & 0.11 & 1004.66 \\
& 0.6 & 2.8228 & 17.9673 & 17.0210 & 5.19 & 7.51 & 1172.72 & 29.4 & 0.57 & 0.15 & 1099.66 \\
& 0.5 & 3.0493 & 20.6382 & 18.2654 & 6.87 & 10.04 & 1914.41 & 39.8 & 0.59 & 0.18 & 1088.88 \\
\hline

\multirow{6}{*}{$s_B=2,\;Y_l=0.2$}
& 1.0 & 2.1336 & 13.3157 & 13.7612 & 0.00 & \ldots & 0.00 & 0.0 & 0.38 & \ldots & \ldots \\
& 0.9 & 2.2341 & 13.9883 & 14.3350 & 1.49 & 3.73 & 151.36 & 4.7 & 0.41 & 0.03 & 637.07 \\
& 0.8 & 2.4057 & 15.2764 & 14.9787 & 2.57 & 4.93 & 425.89 & 12.8 & 0.44 & 0.07 & 828.63 \\
& 0.7 & 2.5415 & 16.0233 & 15.7908 & 3.56 & 5.51 & 658.07 & 19.1 & 0.51 & 0.11 & 1028.75 \\
& 0.6 & 2.7600 & 17.5571 & 16.7340 & 4.96 & 7.01 & 1093.34 & 29.4 & 0.57 & 0.15 & 1126.56 \\
& 0.5 & 2.9812 & 20.1458 & 18.0776 & 6.55 & 9.33 & 1771.42 & 39.7 & 0.60 & 0.18 & 1118.44 \\
\hline

\multirow{6}{*}{$s_B=2,\;Y_{\nu_e}=0$}
& 1.0 & 2.0861 & 13.1228 & 13.6825 & 0.00 & \ldots & 0.00 & 0.0 & 0.37 & \ldots & \ldots \\
& 0.9 & 2.1925 & 13.7832 & 14.2377 & 1.44 & 3.54 & 145.48 & 5.1 & 0.41 & 0.03 & 647.47 \\
& 0.8 & 2.3236 & 14.6884 & 14.8915 & 2.36 & 4.25 & 345.57 & 11.4 & 0.45 & 0.07 & 884.05 \\
& 0.7 & 2.4931 & 15.7841 & 15.7026 & 3.43 & 5.22 & 623.82 & 19.5 & 0.50 & 0.10 & 1045.35 \\
& 0.6 & 2.7052 & 17.3847 & 16.6721 & 4.77 & 6.71 & 1050.47 & 29.7 & 0.56 & 0.14 & 1132.52 \\
& 0.5 & 2.9129 & 19.9439 & 17.9921 & 6.26 & 8.87 & 1685.48 & 39.6 & 0.58 & 0.18 & 1124.19 \\
\hline

\multirow{6}{*}{$T=0$}
& 1.0 & 2.0467 & 12.8948 & 13.5125 & 0.00 & \ldots & 0.00 & 0.0 & 0.37 & \ldots & \ldots \\
& 0.9 & 2.1506 & 13.5423 & 14.0566 & 1.38 & 3.34 & 137.24 & 5.1 & 0.41 & 0.03 & 658.52 \\
& 0.8 & 2.2791 & 14.4276 & 14.7313 & 2.27 & 4.01 & 325.62 & 11.4 & 0.45 & 0.07 & 899.94 \\
& 0.7 & 2.4445 & 15.5020 & 15.5128 & 3.29 & 4.92 & 586.97 & 19.4 & 0.50 & 0.10 & 1064.36 \\
& 0.6 & 2.6521 & 17.0699 & 16.5186 & 4.58 & 6.31 & 986.30 & 29.6 & 0.56 & 0.14 & 1154.08 \\
& 0.5 & 2.8527 & 19.5715 & 17.7908 & 5.98 & 8.29 & 1565.58 & 39.4 & 0.58 & 0.18 & 1147.75 \\
\hline\hline
\end{tabular}
\end{adjustbox}
\label{Mass_tab1}
\end{table*}

The two DDQM parameterizations adopted from the Bayesian analysis~\cite{daSilva:2023okq} produce distinct equilibrium sequences, as summarized in \cref{Mass_tab1} and \cref{Mass_tab}. The stiffer set ($C=0.8,\ \sqrt{D}=127.4\ \text{MeV}$) yields systematically larger maximum masses, radii, angular momenta, moments of inertia, and quadrupole moments than the softer set ($C=0.65,\ \sqrt{D}=133.2\ \text{MeV}$), at every evolutionary stage and rotation rate. For example, at the Keplerian limit ($r_p/r_e=0.5$) in the cold catalyzed stage, \cref{Mass_tab} gives $M_{\max}=3.05\,M_\odot$, $R(M_{\max})=21.33\ \text{km}$, $I=10.47\times10^{45}\ \text{g cm}^2$, and $|Q|=1985\times10^{42}\ \text{g cm}^3$, compared to $2.85\,M_\odot$, $19.57\ \text{km}$, $8.29\times10^{45}\ \text{g cm}^2$, and $1566\times10^{42}\ \text{g cm}^3$ in \cref{Mass_tab1}. Despite these differences, the fractional increase in maximum mass due to rotation, $\Delta M_{\max}(\%)$, is nearly identical between the two parameterizations (about $40\%$ for hot, lepton-rich stages and $39\%$ for cold stars). The softer EOS (lower maximum mass, more easily deformable) allows higher spin frequencies at the Keplerian limit across all evolutionary stages. These comparisons illustrate that the DDQM model’s predictions span a measurable range, and future multi-messenger observations (e.g., simultaneous $M$--$R$--$I$ constraints) can help discriminate between the two parameterizations and thus better constrain the underlying QM EOS.

 Using \cref{Mass_tab} as a case study, increasing rotation (decreasing $r_p/r_e$) systematically raises the maximum mass ($M$), equatorial radius ($R_e$), angular momentum ($J$), moment of inertia ($I$), and quadrupole moment ($Q$), while the ratio $T_{\rm kin}/|W|$ increases to $\sim 0.18$--$0.19$ near the Keplerian limit ($\nu_K$), indicating strong centrifugal support. The spin frequency ($\nu$) peaks for the most oblate configurations, and centrifugal flattening leads to enhanced $I$ and quadrupole deformations. Along the $M$ sequence, the polar redshift $Z_p$ also increases; the increase in mass and decrease in polar radius $r_p$ together outweigh the centrifugal weakening of the gravitational potential.

At fixed rotation, thermal evolution introduces a clear and systematic ordering: as the star deleptonizes from hot, lepton-rich to cold, catalyzed states, the $M$ decreases slightly, $R_e$ contracts, and the $J$, $I$, and $Q$ all decline as the star becomes more compact. The temperature $T$ at a fixed mass of $2.0\,\rm M_\odot$ shows two competing effects: it decreases with increasing rotation (rotational cooling), but increases with $s_B$ during deleptonization before vanishing in the cold, catalyzed limit. In contrast, the $\nu$ at $M$ rises with cooling, since a colder, stiffer EOS allows faster rotation before mass shedding, while $T_{\rm kin}/|W|$ remains primarily determined by the degree of flattening rather than temperature. 

{The relative increase in the $M_{\rm max}$, grows monotonically with rotation and reaches values of up to $\sim 40\%$ at the Kepler limit for hot, lepton-rich configurations. This enhancement exceeds the typical $\sim 20\%$ found for hadronic stars \cite{Gourgoulhon:1999vx, Zhou:2019hyy, Gondek-Rosinska:2000rjg} and is consistent with universal relations predicting $\sim 15\!-\!20\%$ for hadronic matter \cite{Konstantinou:2022vkr} and $\sim 20\!-\!33\%$ for self-bound QSs \cite{Konstantinou_2026}. A direct comparison with Ref.~\cite{daSilva:2025cfe}, which studied rotating stars composed of nucleonic, hyperonic, and $\Delta$-baryon matter, shows that purely nucleonic EOS yield a $\sim 9\!-\!13\%$ increase at the Keplerian limit, while the inclusion of $\Delta$-isobars leads to enhancements of up to $\sim 21\%$.}

The large values of $T_{\rm kin}/|W|$ attained near the $\nu_K$~($\sim 0.18$--$0.19$) indicate that rapidly rotating proto-QSs are susceptible to the Chandrasekhar--Friedman--Schutz secular instability, which can drive sustained gravitational-wave emission from non-axisymmetric modes~\cite{Stergioulas:2003yp}. The softer EOS and enhanced equatorial expansion of QM allow these stars to reach higher $T_{\rm kin}/|W|$ than their hadronic counterparts, strengthening their prospects as detectable sources of continuous gravitational waves from young compact remnants~\cite{Andersson:2002ch, Gondek-Rosinska:2001kuc}. Astrophysically, these trends imply that rapidly rotating, hot proto-QSs can support very large masses and develop strong quadrupole deformations, making them promising gravitational-wave sources and potentially long-lived against collapse. They further highlight that mass, radius, spin, redshift, and gravitational-wave information must be combined to reliably infer the underlying EOS and thermal state, as rotation and temperature effects can otherwise mask the true nature of the stellar interior.

\begin{figure}[!t]	 		
  \includegraphics[width=0.5\textwidth]{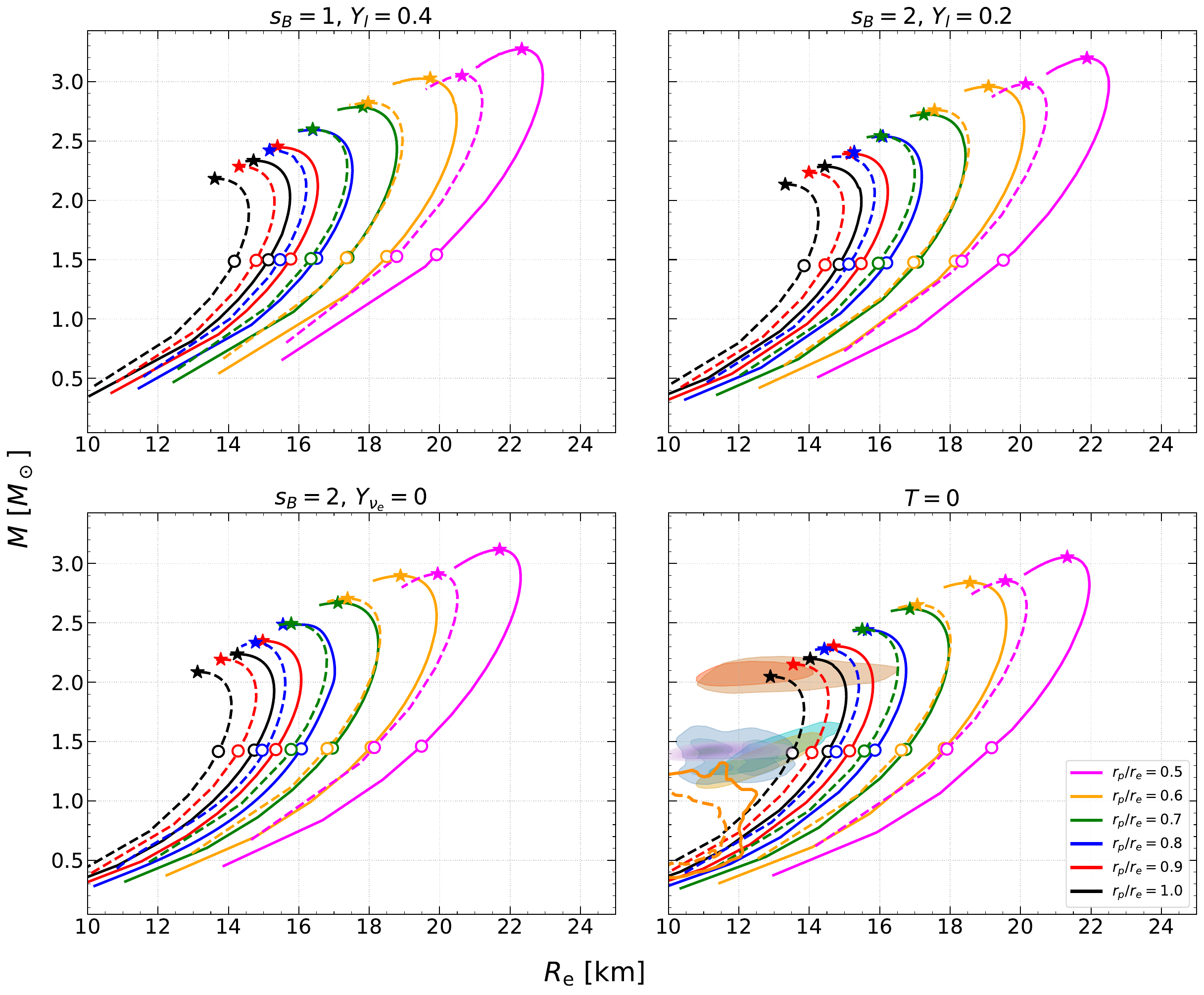}
			 			\caption{The gravitational mass $M$ of rotating proto-QSs as a function of equatorial radius $R_e$. The solid lines correspond to EOS prameterization $C = 0.8, \,\sqrt{D}=127.4$\,MeV and the dashed lines correspond to $C = 0.65, \,\sqrt{D}=132.2$\,MeV~\cite{daSilva:2023okq}. In all four panels, the small circles along each curve represent the structural evolution of a single star with a fixed baryonic mass of \(M_b = 1.55\,M_\odot\). This serves as the baseline configuration, corresponding to \(\sim 1.45\, M_\odot\) cold, catalyzed QSs. The gravitational mass varies with the evolutionary stages. In the $T=0$ panel, we also have observational confidence contours: steel blue indicates the binary components of GW170817, with their respective 90\% and 50\% credible intervals. The plot includes the 1 $\sigma$ (68\%) CI for the 2D mass-radius posterior distributions of the millisecond pulsars PSR J0030 + 0451 (in cyan and yellow color) \cite{riley2019, Miller:2019cac} and PSR J0740 + 6620 (in orange and peru color)\cite{riley2021, Miller:2021qha}. Furthermore, we display the latest NICER measurements for the mass and radius of PSR J0437-4715 \cite{Choudhury:2024xbk} (lilac color). The supernova remnant HESS J1731$-$347 \cite{2022NatAs...6.1444D} is shown in red, with the outer contour representing the 90\% CL and the inner contour representing the 50\% CL.}
		\label{Mass_Rep}	 	
     \end{figure}

The gravitational mass–equatorial radius sequences shown in \cref{Mass_Rep} demonstrate how finite-temperature effects and rotation govern the structural evolution of proto-QSs, with hot, lepton-rich configurations exhibiting larger radii than the cold, catalyzed limit. Similar trends were found for nonrotating proto-QSs in \cite{Issifu:2023qoo, Kumari:2021tik}. This behavior contrasts with purely hadronic protoneutron stars, which are more compact and support lower maximum masses in the lepton-rich phase, then expand and reach higher maximum masses during the hotter, higher-entropy and lepton-poor deleptonization stage, before contracting again in the cold limit \cite{Pons:1998mm, Prakash:1996xs, Issifu:2023qyi}. The contrast reflects the underlying microphysics: hadronic matter stiffens during deleptonization due to thermal pressure and repulsive nuclear interactions, whereas QM is already thermally inflated at early times and contracts monotonically as thermal support is removed during cooling.

As rotation increases, quantified by a decreasing $r_p/r_e$, centrifugal support enlarges the $R_e$ and raises the maximum gravitational mass, as indicated by the star symbols along each sequence. The most oblate configurations ($r_p/r_e \approx 0.5$) approach the Keplerian limit (see \cref{Mass_Ref}). This qualitative behavior is consistent with that of hadronic stars \cite{daSilva:2025cfe}(and references therein); however, for uniform rotation, QSs are expected to enhance their maximum mass by up to $\sim40\%$ \cite{Gondek-Rosinska:2000rjg, Gourgoulhon:1999vx}, compared to $\sim20\%$ for hadronic stars, with even larger increases anticipated for differential rotation \cite{Szkudlarek:2019odl, Zhou:2019hyy}, {where the angular velocity varies with position}. This is attributed to QM’s softer high-density response and larger equatorial expansion under rotation. Consequently, rapidly rotating proto-QSs occupy a distinct region of the mass--radius plane, where both the enlarged radius and the rotation-induced mass enhancement must be accounted for when constraining the underlying EOS.

In the final panel, corresponding to cold and catalyzed configurations, QS matter with the parameterization \(C = 0.65\) and \(\sqrt{D} = 132.2\,\text{MeV}\) (dashed lines) is consistent with all the observational constraints shown in the figure for the static configuration \((r_p/r_e = 1)\), and slowly rotating case \((r_p/r_e = 0.9)\) also satisfies most of the constraints. In contrast, the parameter set \(C = 0.8\) and \(\sqrt{D} = 127.4\,\text{MeV}\), which produces higher maximum mass (solid lines), satisfies the constraints from HESS~J1731--347~\cite{Horvath:2023uwl} and NICER PSR~J0740+6620~\cite{Miller:2021qha} in the static limit while the slow rotation \((r_p/r_e = 0.9)\) statisfies PSR~J0740+6620. This implies that the structure of the stars is dependent on the EOS parameterization. We emphasize that the observed pulsar contours in \cref{Mass_Rep} are used strictly as empirical constraints, not as evidence for QSs. Rather, they delineate the region of parameter space that any viable EOS must reproduce. 

\begin{figure*}[!t]	 		
  \includegraphics[width=\textwidth]{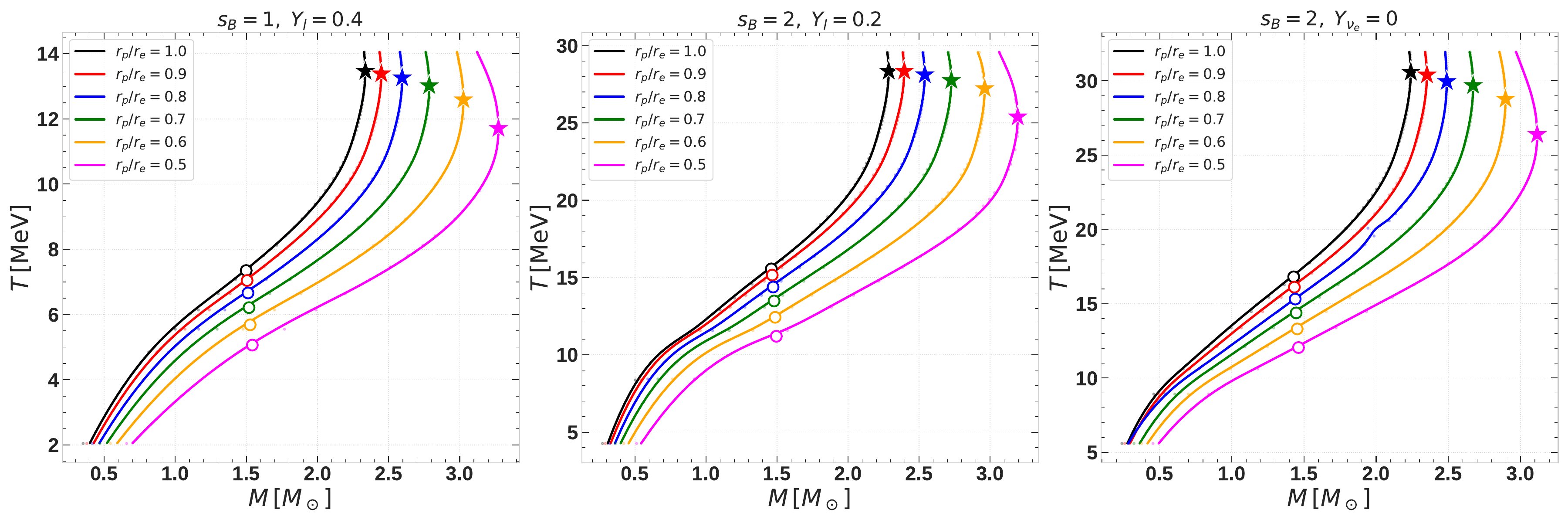}
			 			\caption{The core temperature evolution of rotating proto-QSs examined along gravitational-mass sequences. The stars in the plots indicate the maximum mass for each configuration. The small open circles in the curve show the thermal evolution of a single star with fixed baryonic mass of 1.55 $\rm M_\odot$.
                        }
		\label{MassxT}	 	
     \end{figure*}
\Cref{MassxT} shows the core temperature as a function of gravitational mass for sequences of rotating proto-QSs. The main result is the substantial influence of rotation, parameterized by ${r_p}/{r_e}$, on both the thermal properties and the stability limits of these stars. Our analysis shows that $Y_l$ mainly governs the thermal state: at fixed $s_B$, neutrino trapping in the early lepton-rich stage increases the core pressure and lowers temperature compared to the later neutrino-transparent stage $(s_B=2,\,Y_l=0)$. Therefore, temperature variations from changes in $(C,\,D)$ are minor, so the thermal structure is set by thermodynamics rather than the EOS parametrization. We therefore show core temperature evolution for a single model: $C=0.8$ and $\sqrt{D}=127.4$ MeV. 

For a fixed stellar mass, increased rotational flattening (lower ${r_p}/{r_e}$) reduces gravitational compression, resulting in systematically lower core temperatures along each sequence. The starred points indicate the maximum stable mass for each configuration, showing that rapid rotation significantly increases the mass threshold for gravitational collapse compared to the non-rotating case. {Along the stable branch, the core temperature rises monotonically with gravitational mass due to increasing central compression, while beyond the maximum mass, the configurations become unstable and the temperature increases as the gravitational mass decreases, signaling a loss of hydrostatic support.} The sequence of panels, from $s_B = 1,\, Y_l = 0.4$ characteristic of the immediate post-formation phase, to the states, $ s = 2,\, Y_{l} = 0.2$ and $ s = 2,\, Y_{\nu_e} = 0$, captures the star’s evolution through deleptonization and neutrino-transparent states. Together, these results map the thermodynamic trajectories accessible to hot, rotating QM and place important constraints on the stability and early evolution of proto-QSs, whose EOS differs fundamentally from that of hadronic stars.

Comparing the temperatures obtained from QM \cite{Issifu:2023qoo} and hadronic EOSs at fixed $s_B$, we find that hadronic matter \cite{Pons:1998mm} systematically attains higher temperatures. This difference originates from the larger effective degeneracy of QM, which includes color and spin degrees of freedom \cite{Lugones:2002zd}, whereas typical hadronic models account primarily for spin degeneracy \cite{Prakash:1996xs}. For a fixed thermodynamic state, the increased number of degrees of freedom in QM distributes the thermal energy among more microscopic modes, resulting in a lower temperature. In the present model, the temperature difference can reach up to $\sim$40\%, suggesting that thermal profiles of newly born compact stars may provide a potential observational discriminator between hadronic and QM compositions.

\begin{figure}[!t]	 		
  \includegraphics[width=0.5\textwidth]{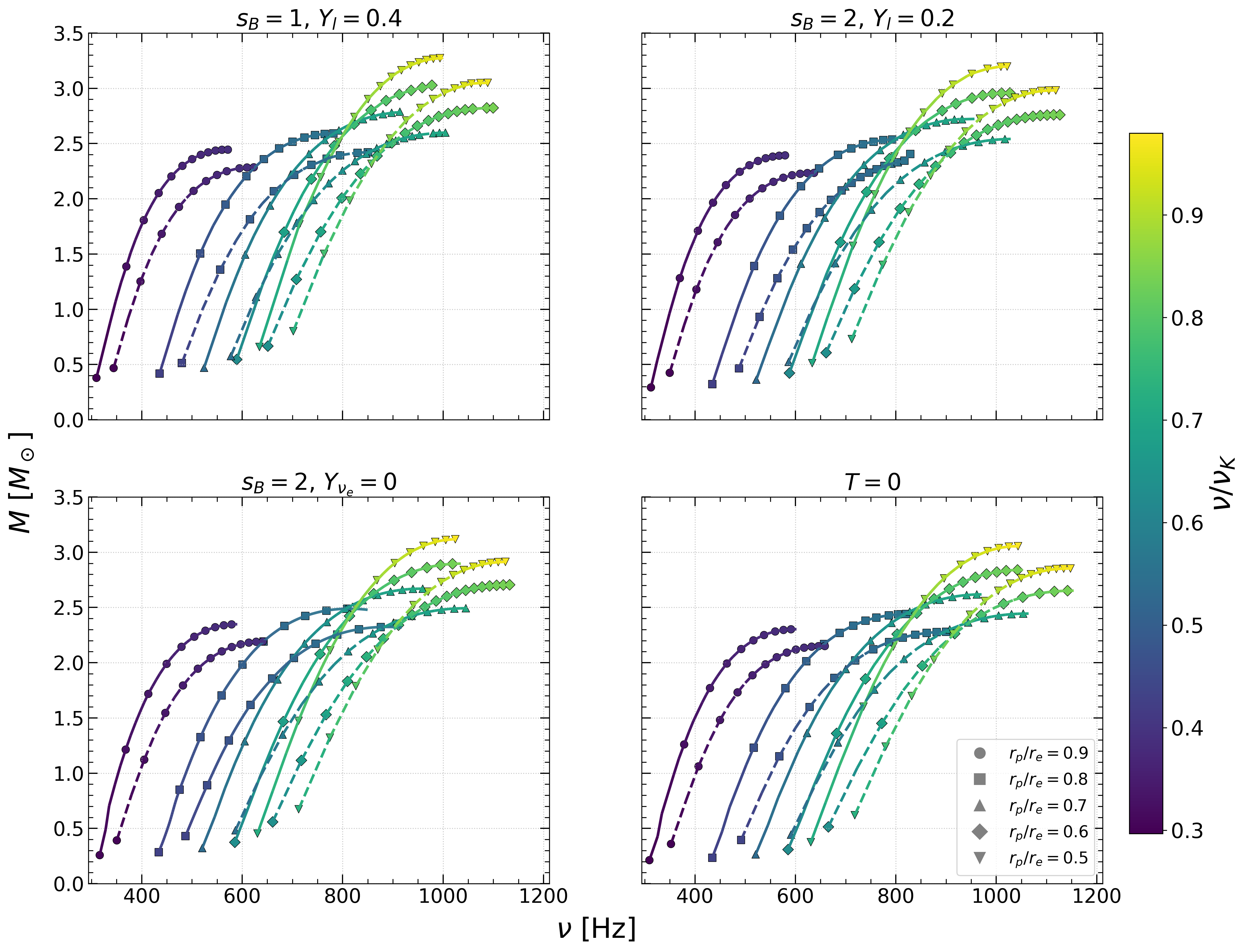}
			 			\caption{Variation of the ratio of the rotational frequency to the Keplerian frequency $\nu/\nu_K$ as a function of the stellar mass. The line styles reflect the different models employed: solid lines represent $C=0.8$ and $\sqrt{D}=127.4$\,MeV while dashed lines represent $C=0.65$ and $\sqrt{D}=133.2$\,MeV.
                        }
		\label{Mass_Ref}	 	
     \end{figure}

The variation of $\nu$ and $\nu/\nu_K$ with $M$ for rotating proto-QSs in \cref{Mass_Ref} highlights how the microscopic physics of deconfined QM governs their rotational behavior across evolutionary stages. The differences between the two models are reflected in the $\nu$ and $\nu/\nu_K$ variations: the $(C=0.8,\, \sqrt{D}=127.4\,\text{MeV})$ parameterization supports lower frequencies at a fixed mass than $(C=0.65,\, \sqrt{D}=133.2\,\text{MeV})$, which exhibits a lower maximum mass. This implies that $(C=0.65,\, \sqrt{D}=133.2\,\text{MeV})$ is more easily deformable, corresponding to a lower $\nu_K$. However, the qualitative trends in both models with respect to the rotation parameter $r_p/r_e$ and the thermodynamic conditions remain the same. 

The ratio $\nu/\nu_{K}$, represented by the color bar in \cref{Mass_Ref}, characterizes how close a star is to the Keplerian limit. Although this quantity is not directly observable, it can be inferred from combined measurements of the stellar mass $M$ and radius $R$, which set the $\nu_K$ through the scaling $\nu_{K} \propto \sqrt{M/R^{3}}$. In practice, estimates of $M$ and $R$ obtained from X-ray pulse profile modeling, such as those provided by NICER \cite{riley2019, Miller:2019cac, riley2021, miller2021} or future missions like eXTP \cite{watts2019dense, eXTP:2018anb}, can be combined with independently measured $\nu$ from radio timing to infer $\nu/\nu_{K}$. Additionally, the physical requirement $\nu \leq \nu_{K}$ provides an upper bound, implying that the highest observed spin frequencies in a population can be used to constrain typical values of this ratio.

In the hot, lepton-rich phases, pressure from thermal quarks and trapped neutrinos softens the EOS \cite{Issifu:2023qoo, Kumari:2021tik}, reducing compactness and lowering the Keplerian frequency $\nu_K$. As a result, $\nu$ and $\nu/\nu_K$ increase rapidly with mass, indicating that hot proto-QSs approach the Keplerian limit at relatively low masses, especially for strongly oblate configurations ($r_p/r_e = 0.5$). This observation is qualitatively similar to what is observed in rotating protoneutron stars using hadronic EOS~\cite{daSilva:2025cfe}. For a general overview of relativistic rotating stars, see Ref.~\cite{Paschalidis:2016vmz}.

As the star cools and deleptonizes, thermal support is lost, and the zero-temperature QM EOS dominates, stiffened by Pauli blocking and repulsive vector interactions \cite{Alford:2004zr}. This increases compactness, raises both the maximum mass and $\nu_K$ \cite{Gondek-Rosinska:2000rjg}, and allows cold QSs to rotate at higher absolute frequencies while remaining farther from the Keplerian limit. Rotation amplifies these trends through centrifugal flattening: hot QM is more susceptible to mass shedding, whereas cold, stiff QM sustains rapid rotation over a wider mass range. Together, these trends encode how quark-level thermodynamics translate into macroscopic rotational stability, offering a potential observational signature to distinguish QSs from hadronic ones~\cite{Weber:2004kj, Oertel:2016bki}.

\begin{figure}[!t]	 		
  \includegraphics[width=0.5\textwidth]{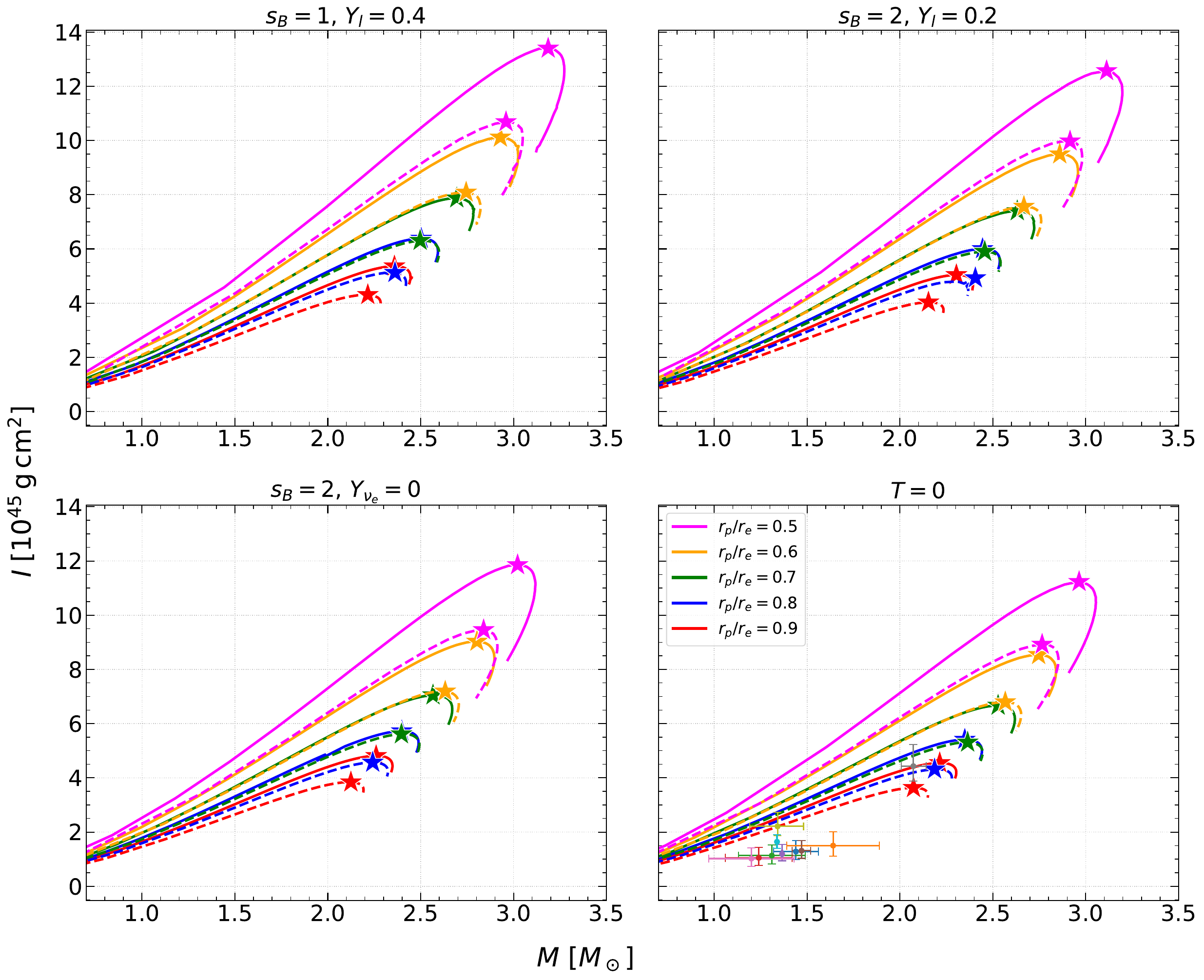}
			 			\caption{The moment of inertia $I$ as a function of the stellar mass for different rotational fattening. The line styles reflect the different models employed: solid lines represent $C=0.8$ and $\sqrt{D}=127.4$\,MeV while dashed lines represent $C=0.65$ and $\sqrt{D}=133.2$\,MeV. Overlaid error bars in panel $T=0$ represent observational constraints from the following pulsars:  
Red: PSR J0437$-$4715,  
Blue: PSR J0751+1807,  
Green: PSR J1713+0747,  
Orange: PSR J1802$-$2124,  
Purple: PSR J1807$-$2500B,  
Brown: PSR J1909$-$3744,  
Pink: PSR J2222$-$0137,  
Gray: PSR J0740+6620 (NICER),  \cite{Li_2022}
Olive: PSR J0030+0451 (NICER), \cite{Silva:2020acr} 
Cyan: PSR J0737$-$3039A (Double Pulsar) \cite{Kumar:2019xgp,PhysRevD.105.063023,Bejger:2005jy}. {The $I$ increases along the stable mass sequence up to $M_{\rm max}$ (starred points along the curve) and then decreases as the mass decreases along the unstable branch.}
}
		\label{Mass_ReI}	 	
     \end{figure}

     \Cref{Mass_ReI} shows the moment of inertia $(I)$ of rotating proto-QSs increases monotonically with $M$ and is strongly enhanced by rotation, as decreasing $r_p/r_e$ redistributes matter outward and raises $I$, with the most oblate configurations approaching the Keplerian limit \cite{Gondek-Rosinska:2000rjg, Breu:2016ufb}. In panel $T=0$, we have overlaid error bars representing the inferred observational constraints from the following pulsars:  Red: PSR J0437$-$4715,  Blue: PSR J0751+1807,  Green: PSR J1713+0747,  Orange: PSR J1802$-$2124,  Purple: PSR J1807$-$2500B,  Brown: PSR J1909$-$3744,  Pink: PSR J2222$-$0137,  Gray: PSR J0740+6620 (NICER),  \cite{Li_2022}Olive: PSR J0030+0451 (NICER), \cite{Silva:2020acr} Cyan: PSR J0737$-$3039A (Double Pulsar) \cite{Kumar:2019xgp,PhysRevD.105.063023,Bejger:2005jy}. The two DDQM parameter sets produce markedly different $I$ predictions, as shown in the $T=0$ panel of \cref{Mass_ReI}. The stiffer set ($C=0.8,\, D=127.4\,\text{MeV}$) yields larger $I$ at a given mass, while the softer set ($C=0.65, D=133.2\,\text{MeV}$) gives lower $I$ and intersects the inferred values for both PSR~J0437-4715 and PSR~J0740+6620. These results demonstrate that the DDQM model can accommodate the range of pulsar moments of inertia inferred from observations, emphasizing the need to consider the full parameter space when one intends to compare with observational data.

     Thermal evolution shifts the $I(M)$ sequences systematically: hot, lepton-rich stages yield the largest $I$ due to a softened QM EOS and larger radii, while deleptonization stiffens the EOS, contracts the star, and reduces $I$ toward the cold, catalyzed limit. These trends encode the temperature and density-dependent strong interactions in QM and show that robust EOS constraints from $I$ measurements must incorporate both rotational deformation and thermal history, with combined $M$--$R$--$I$ data to provide a consistent discriminator between quark and hadronic interiors.

     Proto-QSs and hadronic protoneutron stars exhibit qualitatively distinct $I$ behavior due to their contrasting EOS. Although $I$ increases with mass and rotation in both cases, hadronic matter remains stiffer at high density, yielding more compact stars with smaller radii and systematically lower $I$ at fixed mass and flattening \cite{1994ApJ...424..846R}. During deleptonization, hadronic stars may undergo a temporary stiffening that produces non-monotonic $I(M)$, (this non-monotonic behavior with thermodynamic conditions can be seen in Fig.~4 of \cite{daSilva:2025cfe}) whereas QM softens in hot, lepton-rich stages, leading to larger radii and moments of inertia that can be \(20\text{--}40\%\) higher for the same mass and spin. Consequently, combined $M$--$R$--$I$ measurements provide a promising discriminator between hadronic and quark stars interiors \cite{Pons:1998mm, Prakash:1996xs, Oertel:2016bki}. 
     
     The trends for both models are qualitatively identical as established in \cref{Mass_Rep}, \cref{Mass_Ref}, and \cref{Mass_ReI}. Since our goal is to study the rotational properties of the proto-QSs hereafter, we focus on the stiffer parameterization ($C=0.8, D=127.4\,\text{MeV}$), which amplifies these properties. All discussions hereafter use the stiffer model, except when comparison with observational data requires the softer parameterization. In such instances, reference will be made explicitly to ($C=0.65, \, D=133.2\,\text{MeV}$) parameterization set.

\begin{figure}[!t]	 		
  \includegraphics[width=0.5\textwidth]{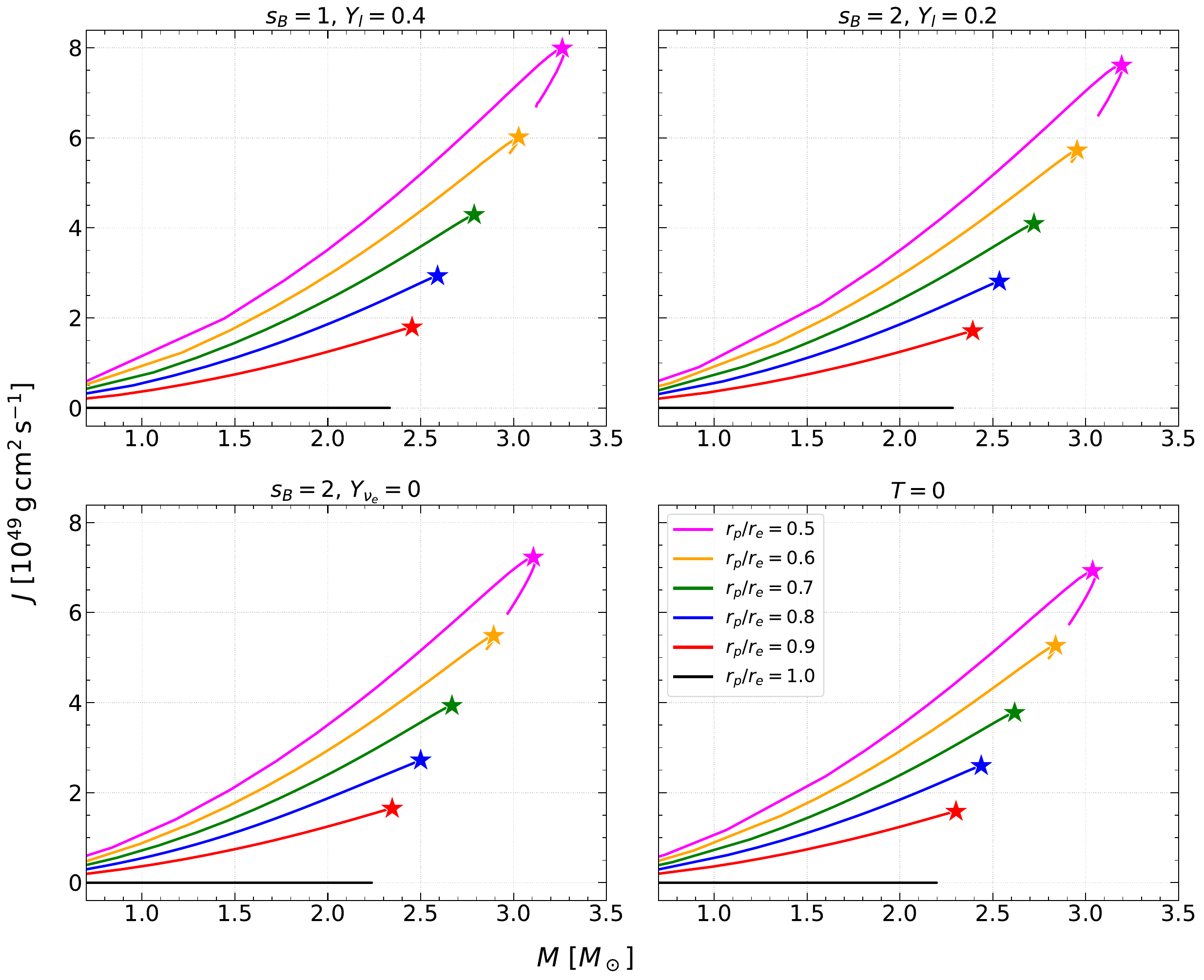}
			 			\caption{Variation of the angular momentum, $J$, as a function of the stellar mass for different rotational flattening.}
		\label{Mass_Rej}	 	
     \end{figure}
\Cref{Mass_Rej} shows that the angular momentum $J$ of rotating proto-QSs increases monotonically with gravitational mass $M$ and rises strongly with rotation, quantified by the decrease of
$r_p/r_e$. Generally, hot, lepton-rich configurations store significantly more angular momentum than the lepton-poor ones \cite{daSilva:2025cfe} because thermal and neutrino pressures stiffen the QM EOS, enhancing its maximum stellar mass, and increasing the moment of inertia (see the discussion in figs.~\ref{Mass_Rep} and \ref{Mass_ReI}). As the star deleptonizes, the EOS stiffens, the star contracts, and the $J(M)$ curves shift downward, with the cold, catalyzed sequences being the most compact. The most oblate configurations ($r_p/r_e \simeq 0.5$) approach the Keplerian mass-shedding limit (see \cref{Mass_Ref}), which sets an upper bound on the attainable angular momentum \cite{Stergioulas:2003yp, 1994ApJ...424..823C}. Overall, the figure highlights that both thermal evolution and rotational deformation critically control the angular-momentum budget of proto-QSs, with newly born lepton-rich objects capable of storing substantially more rotational energy than their cold lepton-poor descendants. 

Comparatively, hadronic protoneutron stars show a similar qualitative increase of angular momentum with mass and rotation, but their stiffer hadronic EOS makes them more compact and able to store less angular momentum than QM stars of equal mass at the same rotational frequency, especially in hot, lepton-rich phases \cite{Oertel:2016bki, daSilva:2025cfe}. Moreover, during deleptonization, hadronic stars may undergo a temporary stiffening that produces a non-monotonic $J(M)$ evolution, absent in QM sequences (see the comparative discussion in \cref{Mass_Ref}), so the combined $M$--$R$--$J$ measurements could discriminate between hadronic and quark stars interior~\cite{Yagi:2016bkt, Silva:2020acr}.

\begin{figure}[!t]	 		
  \includegraphics[width=0.5\textwidth]{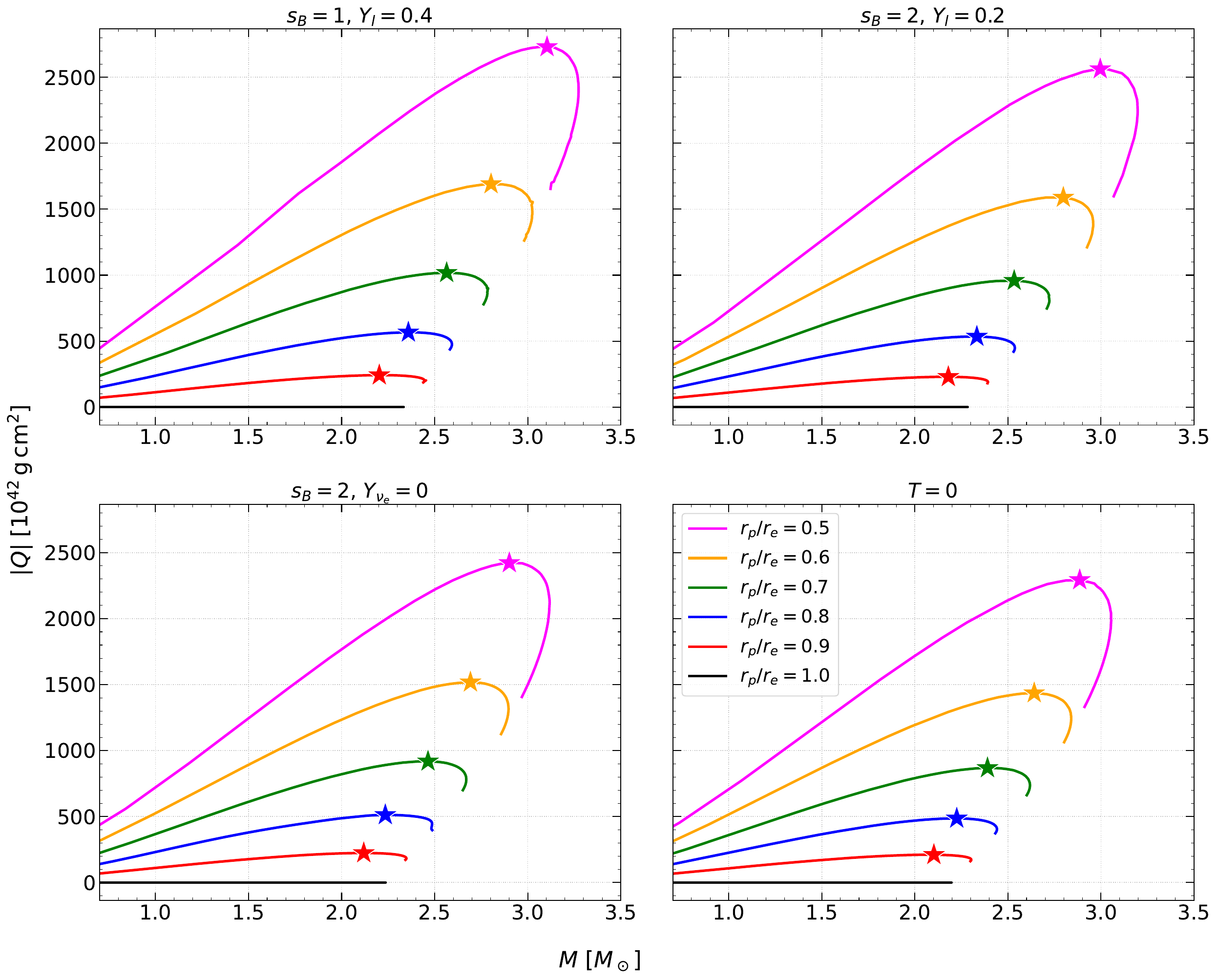}
			 			\caption{Variation of the magnitude of the quadrupole moment $|Q|$ as a function of gravitational mass $M$ for different rotational flattening.}
		\label{Mass_Req}	 	
     \end{figure}

The $|Q|(M)$ sequences in \cref{Mass_Req} show that the magnitude of the quadrupole moment ($|Q|$) of rotating proto-QSs increases monotonically with $M$ and is strongly amplified by rotation, with smaller $r_p/r_e$ producing the higher deformation through centrifugal flattening \cite{Laarakkers:1997hb}. Thermal evolution shifts these curves systematically: hot, lepton-rich stars exhibit the largest $|Q|$ because a softened QM EOS and larger radii make them more easily deformable \cite{Issifu:2023qoo}, while deleptonization stiffens the EOS, contracts the star, and reduces $|Q|$ toward the cold, catalyzed limit. The $|Q|$ is a fundamental parameter of the external spacetime, influencing effects such as the tidal response of compact objects in binaries, which imprints on the gravitational-wave signal during inspiral, and contributing to the I--Love--Q universal relations. Future multimessenger observations that combine binary pulsar timing, X-ray measurements, and gravitational-wave detections may therefore offer a viable path to distinguish between different quadrupole-moment scenarios \cite{Mashhoon:2006fj}.

The evolution of the $|Q|$ reflects the temperature and density-dependent strong interactions in QM and has implications for gravitational-wave emission, as hot, rapidly rotating proto-QSs are stronger emitters whose signals weaken as the star cools~\cite{Andersson:2002ch, Cutler:2002nw}. Compared to hadronic protoneutron stars, which are more compact and less deformable due to a stiffer high-density EOS, QSs of the same mass and rotation develop larger quadrupole moments, especially in the early hot phases, making combined mass, radius, and quadrupole constraints a promising probe of the stellar interior \cite{Laarakkers:1997hb, Oertel:2016bki}.

\begin{figure}[!t]	 		
  \includegraphics[width=0.5\textwidth]{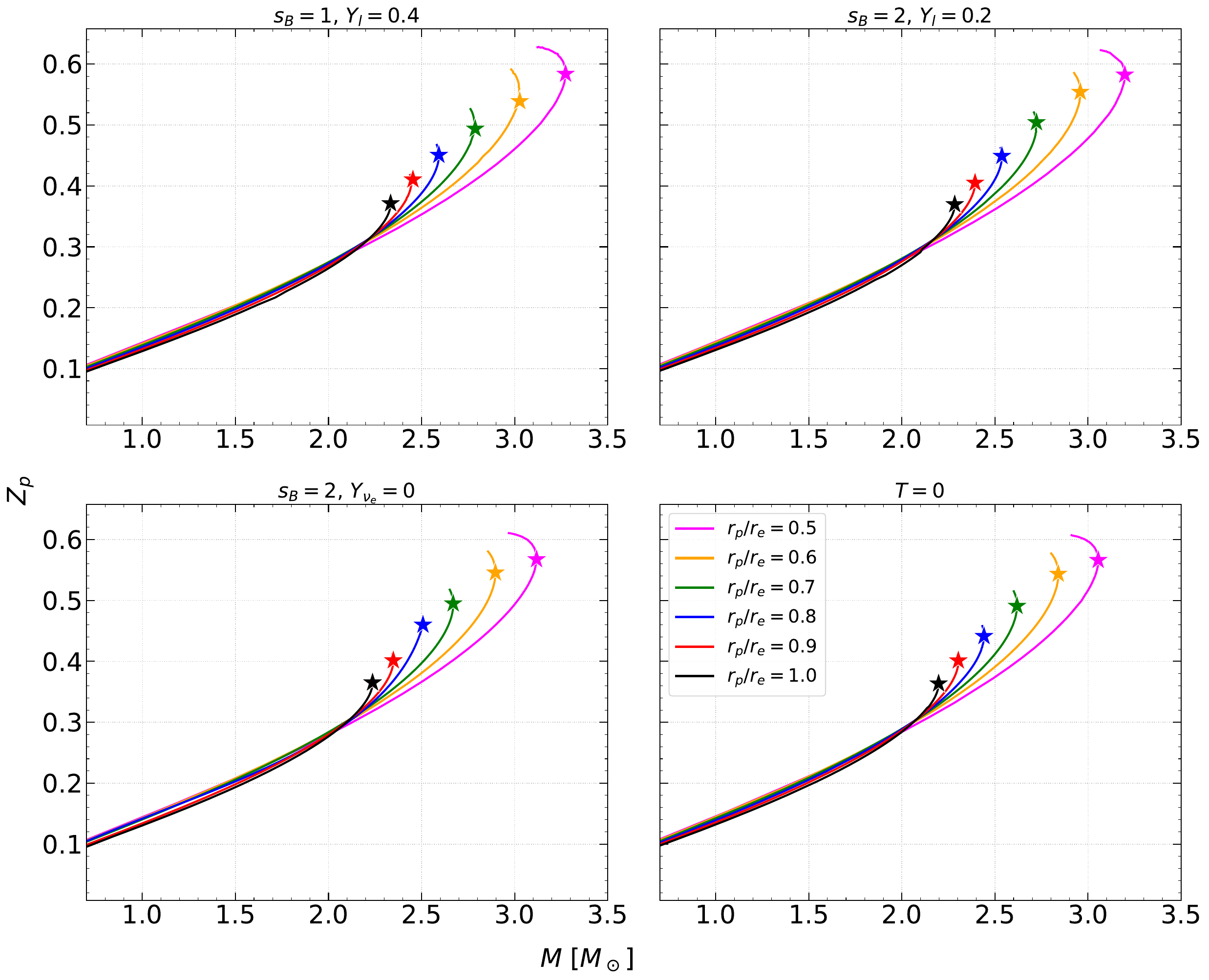}
			 			\caption{Variation of the polar redshift $Z_p$ against the gravitational mass for varying rotational flatness.}
		\label{Mass_Rez}	 	
     \end{figure}

The polar redshift $Z_p$ of rotating proto-QSs shown in \cref{Mass_Rez} increases monotonically with $M$ at all evolutionary stages, reflecting the increase in stellar compactness \cite{Lattimer:2000nx, 1994ApJ...424..823C, Stergioulas95}. Rotation systematically reduces $Z_p$, as centrifugal support lowers the effective gravitational potential, with highly oblate configurations exhibiting the smallest values at fixed mass \cite{Paschalidis:2016vmz}. Thermal evolution introduces an additional ordering: hot, lepton-rich stars tend to exhibit lower $Z_p$, as thermal effects generally lead to larger radii and reduced compactness, whereas cooling and deleptonization contract the star and shift the sequences toward the cold, catalyzed limit where $Z_p$ is maximal. 

Comparisons with hadronic stars indicate that, for a given mass and $r_p/r_e$, hot quark stars may exhibit lower polar redshifts, although this behavior is model dependent and reflects differences in the underlying EOS and thermal response \cite{Pons:1998mm, Gondek-Rosinska:2000rjg}. This highlights that $Z_p$ must be interpreted in conjunction with the rotational state and thermal history, as different configurations may produce similar redshift values. Combined constraints on $Z_p$, $M$, and $R$ can therefore provide a valuable probe of the dense-matter EOS \cite{Yagi:2016bkt}.

\begin{figure}[t!]	 		
  \includegraphics[width=0.5\textwidth]{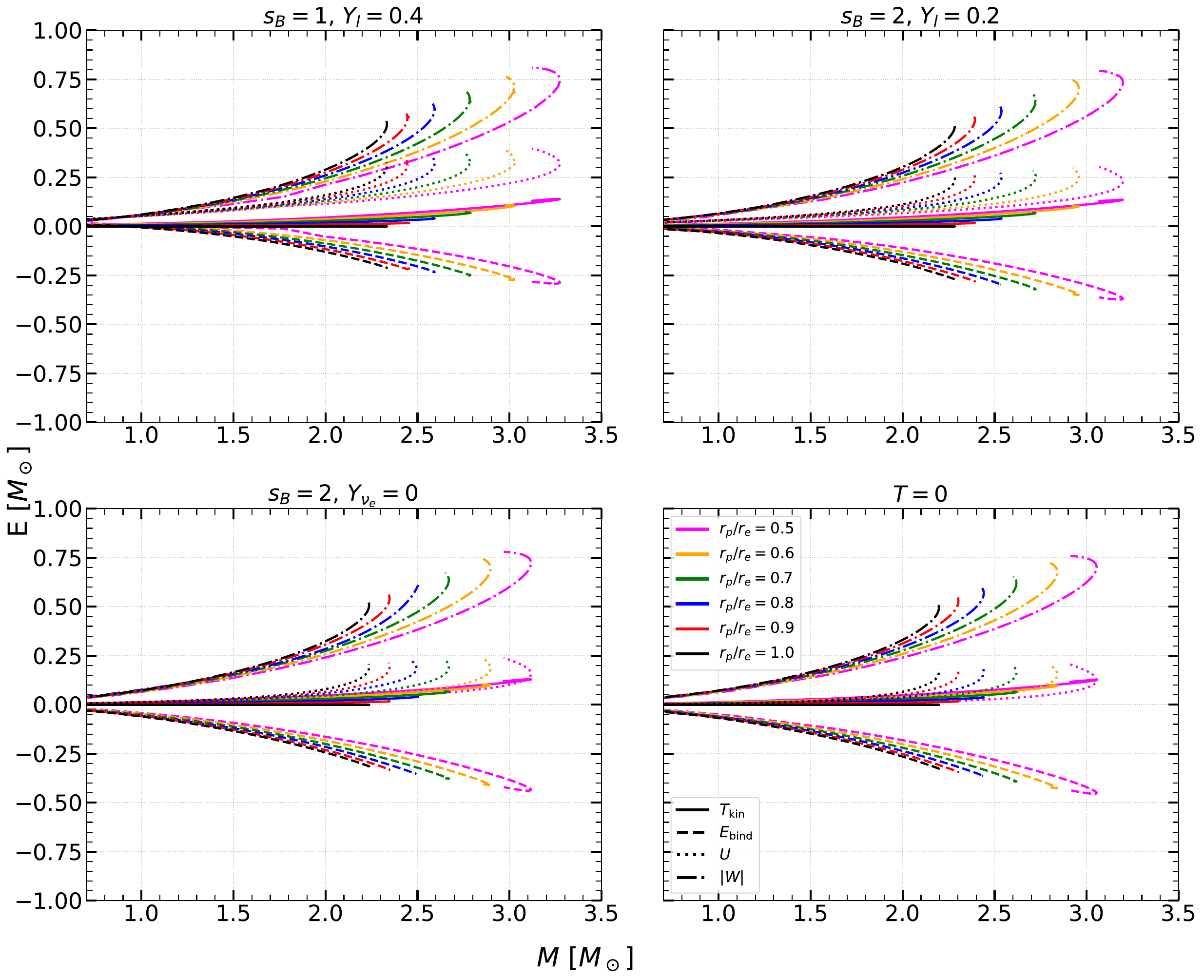}
			 			\caption{The complete energy decomposition of rotating proto-QSs, including the gravitational potential energy $|W|$, rotational kinetic energy $T_{\rm kin}$, binding energy $E_{\rm bind}$, and internal energy, shown as functions of the gravitational mass. }
		\label{MassxT_Ree}	 	
     \end{figure}

\Cref{MassxT_Ree} shows how the gravitational potential energy $|W|$, rotational kinetic energy $T_{\rm{kin}}$, binding energy $E_{\rm{bind}}$, and internal energy $U$ of rotating proto-QSs vary with $M$ across their evolution from neutrino-trapped through deleptonization to cold, catalyzed states. All components grow in magnitude with mass: $|W|$ deepens as self-gravity strengthens \cite{1994ApJ...424..823C}, $T_{\rm{kin}}$ increases as more mass participates in rotation \cite{Stergioulas:2003yp}, $E_{\rm{bind}}$ becomes more negative, and $U$ reflects the amount and thermal content of the QM. Thermal evolution shifts the energy balance: at fixed rotation ($r_p/r_e$), hot, lepton-rich stars have larger internal energy $U$ and a shallower $|W|$, while cooling and deleptonization reduce $U$ and deepen $|W|$ toward the cold, catalyzed limit \cite{Burrows:1986me}.  Rotation further reshapes the balance: decreasing $r_p/r_e$ enhances $T_{\rm{kin}}$ through centrifugal flattening, slightly reduces $|W|$, and weakens the binding, effects that are strongest near the Keplerian limit. These trends reflect the temperature and density-dependent strong interactions in QM and show how composition, thermal state, and spin jointly govern the energy budget and evolution of newly born compact stars. A first comprehensive energy decomposition of rotating QSs was done in \cite{Issifu:2025hqg} for cold rotating QSs.

Compared to hadronic matter, the softer QM EOS in hot, lepton-rich phases yields larger radii and lower compactness, leading to smaller gravitational binding $|W|$ and higher rotational kinetic energy $T_{\rm kin}$ at fixed mass and spin \cite{1996csnp.book.....G, Oertel:2016bki}. By contrast, the stiffer nuclear EOS of hadronic stars produces more compact configurations with deeper gravitational potentials, resulting in larger $|W|$ and a reduced ability to store rotational energy, particularly at early times. By linking the star’s mass, rotation, and thermal state to distinct energy partitions, this decomposition helps interpret observed spins, cooling rates, and stability limits of young compact stars. In particular, comparing inferred rotational energy, binding, and thermal content from timing, neutrino, or gravitational-wave data can help distinguish hot proto-QSs from hadronic ones and constrain the underlying EOS \cite{Lattimer:2004pg, Will:2018ont}.

\section{Conclusion}\label{conc}
This work presents the first systematic investigation of rigidly rotating proto-QSs based on isentropic EOS within the DDQM framework. We employ two model parameterization sets obtained from Bayssian inference using current astrophysical data: $C=0.8$, $\sqrt{D}=127.4$\,MeV and $C=0.65$, $\sqrt{D}=133.2$\,MeV to improve the predictive power of the model and its agreement with observational data. By constructing sequences of equilibrium configurations across four distinct evolutionary stages, from hot, lepton-rich matter to cold, catalyzed QM, we quantified the combined impact of thermal evolution and rotation on the global structure and key observables of these objects. 

We emphasize that the two DDQM parametrizations considered in this work are not intended to provide a universal description of all compact stars. While one parametrization reproduces a broader subset of the currently considered observational constraints, including HESS~J1731$-$347, PSR~J0030+0451, PSR~J0740+6620, and GW170817, the other satisfies only a subset of these constraints, particularly HESS~J1731$-$347 and PSR~J0740+6620. Moreover, additional observational candidates, such as the high-mass black widow pulsar PSR~J0952$-$0607 and very low-mass compact objects like XTE~J1810$-$197, may not be simultaneously reproduced within a single static EOS framework \cite{daSilva:2023okq}. This reflects the current uncertainties in the dense-matter EOS and may indicate that the compact-star population contains objects with different microscopic compositions, including hadronic stars, hybrid stars, and quark stars. Therefore, the observational constraints employed here should be interpreted primarily as empirical benchmarks delimiting the viable DDQM parameter space. In this context, our results are consistent with a coexistence scenario in which NSs and QSs populate different regions of the compact-star landscape~\cite{1986ApJ...310..261A, Weber:2004kj, Drago:2015dea}. A detailed discussion of the possible coexistence of hadronic and quark stars was first presented in \cite{Drago:2013fsa}.

Our results show that rotation substantially enhances the maximum stable mass, equatorial radius, angular momentum, moment of inertia, and quadrupole deformation, with mass increases reaching up to $\sim 40\%$ (independent of the model parameterization) at the Keplerian limit (see $\rm M_{max}$ data in Tabs.~\ref{Mass_tab} and \ref{Mass_tab1}). The ratio of rotational kinetic to gravitational binding energy approaches $T_{\rm kin}/|W| \simeq 0.18$--$0.19$, indicating an increased susceptibility to non-axisymmetric instabilities and potential gravitational-wave emission. At fixed rotation, thermal evolution introduces systematic trends: as the stars cool and deleptonize, they become more compact and their global rotational quantities decrease (see figs.~\ref{Mass_Rep} to \ref{Mass_Rez} and the results summary on Tabs.~\ref{Mass_tab} and \ref{Mass_tab1}).

The mass--radius relations demonstrate that hot proto-QSs are significantly larger than their cold counterparts, while rotation shifts the sequences to higher masses and radii. The softer EOS parameterization is consistent with the observational constraints from GW170817, PSR J0030+0451, PSR~J0740+6620, and HESS~J1731--347, whereas the stiffer parameterization agrees only with HESS~J1731--347 and PSR~J0740+6620 (see the dashed curves for the soft configuration and the solid curves for the stiffer configuration in \cref{Mass_Rep}). The thermal and rotational analyses further reveal that softer EOS parameterization at the early stage leads to lower Keplerian frequencies, whereas cooling and stiffening allow colder configurations to sustain faster absolute rotation. Distinctive trends emerge in key rotational observables. The moment of inertia, angular momentum, and quadrupole moment are strongly amplified in hot, rapidly rotating configurations, while the polar redshift is systematically reduced by centrifugal support and thermal expansion. The complete energy decomposition highlights how the balance between gravitational potential, internal, binding, and rotational energies evolves with both temperature and spin, providing complementary insight beyond standard bulk observables (see \cref{MassxT_Ree}).

Overall, our results demonstrate that the rotational properties of proto-QSs differ both qualitatively and quantitatively from those of hadronic protoneutron stars. These differences, encoded in the mass--radius relation, moment of inertia, quadrupole moment, and energy partitioning, offer a set of potential observational discriminants. Future multimessenger observations combining mass, radius, spin, and gravitational-wave information will be essential to test the QM hypothesis within the DDQM with temperature dependence. This work establishes a robust framework for future extensions, including differential rotation, magnetic fields, and dynamical stability studies of newly born compact stars.

\begin{acknowledgments}
A. I. acknowledges financial support from the São Paulo State Research Foundation (FAPESP), Grant Nos. 2023/09545-1 and 2025/17347-0. This work is part of the project INCT-FNA (Proc. No. 464898/2014-5) and is also supported by the National Council for Scientific and Technological Development (CNPq) under Grants No.  306834/2022-7 (T.F.). T. F. also thanks the financial support from  Improvement of Higher Education Personnel CAPES (Finance Code 001) and FAPESP Thematic Grants (2023/13749-1 and 2024/17816-8). P.~Thakur is supported by the National Research Foundation of Korea (NRF) grant funded by the Korea government (MSIT) (No.~RS-2024-00457037). This work was supported (in part) by the Yonsei University Research Fund(Yonsei University Frontier Fellowship for Postdoctoral Researchers) of 2025. A.K. acknowledges financial support from “The three-dimensional structure of the nucleon from lattice
QCD” 3D-N-LQCD program, funded by the University of Cyprus, and from the projects HyperON (VISION ERC - PATH 2/0524/0001) and Baryon8 (POST-DOC/0524/0001), co-financed by the European Regional Development Fund and the Republic of Cyprus through the Research and Innovation Foundation.

\end{acknowledgments}

\bibliography{refrences}

@article{Chen:2023bxx,
    author = "Chen, Kenneth and Lin, Lap-Ming",
    title = "{Fully general relativistic simulations of rapidly rotating quark stars: Oscillation modes and universal relations}",
    eprint = "2307.01598",
    archivePrefix = "arXiv",
    primaryClass = "gr-qc",
    doi = "10.1103/PhysRevD.108.064007",
    journal = "Phys. Rev. D",
    volume = "108",
    number = "6",
    pages = "064007",
    year = "2023"
}

@article{Szkudlarek:2019odl,
    author = {Szkudlarek, Magdalena and Gondek-Rosi{\'n}ska, Dorota and Villain, Lo{\"\i}c and Ansorg, Marcus},
    title = "{Maximum Mass Of Differentially Rotating Strange Quark Stars}",
    eprint = "1904.03759",
    archivePrefix = "arXiv",
    primaryClass = "astro-ph.HE",
    doi = "10.3847/1538-4357/ab1752",
    journal = "Astrophys. J.",
    volume = "879",
    number = "1",
    pages = "44",
    year = "2019"
}

@article{Madsen:1998uh,
    author = "Madsen, Jes",
    editor = "Cleymans, J. and Geyer, H. B. and Scholtz, F. G.",
    title = "{Physics and astrophysics of strange quark matter}",
    eprint = "astro-ph/9809032",
    archivePrefix = "arXiv",
    doi = "10.1007/BFb0107314",
    journal = "Lect. Notes Phys.",
    volume = "516",
    pages = "162--203",
    year = "1999"
}

@article{Horvath:2023uwl,
    author = "Horvath, J. E. and Rocha, L. S. and de S{\'a}, L. M. and Moraes, P. H. R. S. and Bar{\~a}o, L. G. and de Avellar, M. G. B. and Bernardo, A. and Bachega, R. R. A.",
    title = "{A light strange star in the remnant HESS J1731{\ensuremath{-}}347: Minimal consistency checks}",
    eprint = "2303.10264",
    archivePrefix = "arXiv",
    primaryClass = "astro-ph.HE",
    doi = "10.1051/0004-6361/202345885",
    journal = "Astron. Astrophys.",
    volume = "672",
    pages = "L11",
    year = "2023"
}

@article{Miller:2021qha,
    author = "Miller, M. C. and others",
    title = "{The Radius of PSR J0740+6620 from NICER and XMM-Newton Data}",
    eprint = "2105.06979",
    archivePrefix = "arXiv",
    primaryClass = "astro-ph.HE",
    doi = "10.3847/2041-8213/ac089b",
    journal = "Astrophys. J. Lett.",
    volume = "918",
    number = "2",
    pages = "L28",
    year = "2021"
}

@article{Shen:2005vh,
    author = "Shen, Jian-yong and Zhang, Yun and Wang, Bin and Su, Ru-Keng",
    title = "{Slowly rotating proto strange stars in quark mass density- and temperature- dependent model}",
    eprint = "gr-qc/0503015",
    archivePrefix = "arXiv",
    doi = "10.1142/S0217751X05022391",
    journal = "Int. J. Mod. Phys. A",
    volume = "20",
    pages = "7547--7566",
    year = "2005"
}

@article{Wen:2005uf,
    author = "Wen, X. J. and Zhong, X. H. and Peng, G. X. and Shen, P. N. and Ning, P. Z.",
    title = "{Thermodynamics with density and temperature dependent particle masses and properties of bulk strange quark matter and strangelets}",
    eprint = "hep-ph/0506050",
    archivePrefix = "arXiv",
    doi = "10.1103/PhysRevC.72.015204",
    journal = "Phys. Rev. C",
    volume = "72",
    pages = "015204",
    year = "2005"
}

@article{Chen:2021fdj,
    author = "Chen, Huai-Min and Xia, Cheng-Jun and Peng, Guang-Xiong",
    title = "{Strangelets at finite temperature in a baryon density-dependent quark mass model}",
    eprint = "2110.09194",
    archivePrefix = "arXiv",
    primaryClass = "hep-ph",
    doi = "10.1103/PhysRevD.105.014011",
    journal = "Phys. Rev. D",
    volume = "105",
    number = "1",
    pages = "014011",
    year = "2022"
}

@article{Issifu:2023qoo,
    author = "Issifu, Adamu and da Silva, Franciele M. and Menezes, D{\'e}bora P.",
    title = "{Proto-strange quark stars from density-dependent quark mass model}",
    eprint = "2311.12511",
    archivePrefix = "arXiv",
    primaryClass = "nucl-th",
    doi = "10.1140/epjc/s10052-024-12828-0",
    journal = "Eur. Phys. J. C",
    volume = "84",
    number = "5",
    pages = "463",
    year = "2024"
}

@InProceedings{ZhouJPhCS2017,
  author    = {Zhou, Enping},
  booktitle = {Journal of Physics Conference Series},
  title     = {{Properties of relativistically rotating quark stars}},
  year      = {2017},
  month     = jun,
  pages     = {012007},
  series    = {Journal of Physics Conference Series},
  volume    = {861},
  adsurl    = {https://ui.adsabs.harvard.edu/abs/2017JPhCS.861a2007Z},
  doi       = {10.1088/1742-6596/861/1/012007},
  eid       = {012007},
}

@article{Gourgoulhon:1999vx,
    author = "Gourgoulhon, E. and Haensel, P. and Livine, R. and Paluch, E. and Bonazzola, S. and Marck, J. A.",
    title = "{Fast rotation of strange stars}",
    eprint = "astro-ph/9907225",
    archivePrefix = "arXiv",
    journal = "Astron. Astrophys.",
    volume = "349",
    pages = "851",
    year = "1999"
}

@article{Zhou:2019hyy,
    author = "Zhou, Enping and Tsokaros, Antonios and Uryu, Koji and Xu, Renxin and Shibata, Masaru",
    title = "{Differentially rotating strange star in general relativity}",
    eprint = "1902.09361",
    archivePrefix = "arXiv",
    primaryClass = "astro-ph.HE",
    doi = "10.1103/PhysRevD.100.043015",
    journal = "Phys. Rev. D",
    volume = "100",
    number = "4",
    pages = "043015",
    year = "2019"
}

@Article{ZhouAN2017,
  author   = {Zhou, E. and Tsokaros, A. and Rezzolla, L. and Xu, R.},
  journal  = {Astron. Nachrichten},
  title    = {{Maximum mass of axisymmetric rotating quark stars}},
  year     = {2017},
  month    = dec,
  number   = {1044},
  pages    = {1044--1047},
  volume   = {338},
  adsurl   = {https://ui.adsabs.harvard.edu/abs/2017AN....338.1044Z},
  doi      = {10.1002/asna.201713432},
  fjournal = {Astronomische Nachrichten},
}

@article{Gondek-Rosinska:2000rjg,
    author = "Gondek-Rosinska, Dorota and Bulik, Tomasz and Zdunik, Leszek and Gourgoulhon, Eric and Ray, Subharthi and Dey, Jishnu and Dey, Mira",
    title = "{Rotating compact strange stars}",
    eprint = "astro-ph/0007004",
    archivePrefix = "arXiv",
    journal = "Astron. Astrophys.",
    volume = "363",
    pages = "1005",
    year = "2000"
}

@article{Breu:2016ufb,
    author = "Breu, Cosima and Rezzolla, Luciano",
    title = "{Maximum mass, moment of inertia and compactness of relativistic stars}",
    eprint = "1601.06083",
    archivePrefix = "arXiv",
    primaryClass = "gr-qc",
    doi = "10.1093/mnras/stw575",
    journal = "Mon. Not. Roy. Astron. Soc.",
    volume = "459",
    number = "1",
    pages = "646--656",
    year = "2016"
}

@article{Cutler:2002nw,
    author = "Cutler, Curt",
    title = "{Gravitational waves from neutron stars with large toroidal B fields}",
    eprint = "gr-qc/0206051",
    archivePrefix = "arXiv",
    doi = "10.1103/PhysRevD.66.084025",
    journal = "Phys. Rev. D",
    volume = "66",
    pages = "084025",
    year = "2002"
}

@article{Andersson:2002ch,
    author = "Andersson, Nils",
    title = "{Gravitational waves from instabilities in relativistic stars}",
    eprint = "astro-ph/0211057",
    archivePrefix = "arXiv",
    doi = "10.1088/0264-9381/20/7/201",
    journal = "Class. Quant. Grav.",
    volume = "20",
    pages = "R105",
    year = "2003"
}

@article{Laarakkers:1997hb,
    author = "Laarakkers, William G. and Poisson, Eric",
    title = "{Quadrupole moments of rotating neutron stars}",
    eprint = "gr-qc/9709033",
    archivePrefix = "arXiv",
    doi = "10.1086/306732",
    journal = "Astrophys. J.",
    volume = "512",
    pages = "282--287",
    year = "1999"
}

@article{Romanowsky:2000zb,
    author = "Romanowsky, Aaron J. and Kochanek, Christopher S.",
    title = "{Dynamics of stars and globular clusters in m87}",
    eprint = "astro-ph/0008062",
    archivePrefix = "arXiv",
    doi = "10.1086/320947",
    journal = "Astrophys. J.",
    volume = "553",
    pages = "722",
    year = "2001"
}

@Article{ZdunikA&A2000,
   author        = {Zdunik, J. L. and Bulik, T. and Kluźniak, W. and Haensel, P. and Gondek-Rosińska, D.},
  title         = {On the Mass of Moderately Rotating Strange Stars in the MIT Bag Model and LMXBs},
  journal       = {Astronomy and Astrophysics},
  year          = {2000},
  volume        = {359},
  pages         = {143--147},
  month         = jul,
  doi           = {10.48550/arXiv.astro-ph/0004278},
  archivePrefix = {arXiv},
  eprint        = {astro-ph/0004278},
  primaryClass  = {astro-ph},
  adsurl        = {https://ui.adsabs.harvard.edu/abs/2000A&A...359..143Z}
}

@Article{StergioulasA&A1999,
 author        = {Stergioulas, N. and Kluźniak, W. and Bulik, T.},
  title         = {Keplerian Frequencies and Innermost Stable Circular Orbits of Rapidly Rotating Strange Stars},
  journal       = {Astronomy and Astrophysics},
  year          = {1999},
  volume        = {352},
  pages         = {L116--L120},
  month         = dec,
  doi           = {10.48550/arXiv.astro-ph/9909152},
  archivePrefix = {arXiv},
  eprint        = {astro-ph/9909152},
  primaryClass  = {astro-ph},
  adsurl        = {https://ui.adsabs.harvard.edu/abs/1999A&A...352L.116S}
}

@Article{LattimerApJ1990,
  author   = {Lattimer, James M. and Prakash, Madappa and Masak, Dieter and Yahil, Amos},
  journal  = {\apj},
  title    = {{Rapidly Rotating Pulsars and the Equation of State}},
  year     = {1990},
  month    = may,
  pages    = {241},
  volume   = {355},
  adsurl   = {https://ui.adsabs.harvard.edu/abs/1990ApJ...355..241L},
  doi      = {10.1086/168758},
}

@misc{pappas2012multipolemomentsnumericalspacetimes,
      title={Multipole Moments of numerical spacetimes}, 
      author={George Pappas and Theocharis A Apostolatos},
      year={2012},
      eprint={1211.6299},
      archivePrefix={arXiv},
      primaryClass={gr-qc},
      url={https://arxiv.org/abs/1211.6299}, 
}

@ARTICLE{1994ApJ...424..846R,
       author = {{Ravenhall}, D.~G. and {Pethick}, C.~J.},
        title = "{Neutron Star Moments of Inertia}",
      journal = {\apj},
         year = 1994,
        month = apr,
       volume = {424},
        pages = {846},
          doi = {10.1086/173935},
       adsurl = {https://ui.adsabs.harvard.edu/abs/1994ApJ...424..846R}
}

@article{daSilva:2025cfe,
    author = "da Silva, Franciele M. and Issifu, Adamu and Santos, Luis C. N. and Frederico, Tobias and Menezes, D{\'e}bora P.",
    title = "{Hyperons and {\ensuremath{\Delta}}{\textquoteright}s in rotating protoneutron stars: Global properties}",
    eprint = "2504.05495",
    archivePrefix = "arXiv",
    primaryClass = "hep-ph",
    doi = "10.1103/3djv-zbcj",
    journal = "Phys. Rev. D",
    volume = "112",
    number = "2",
    pages = "023007",
    year = "2025"
}

@article{Issifu:2023qyi,
    author = "Issifu, Adamu and Marquez, Kauan D. and Pelicer, Mateus R. and Menezes, D{\'e}bora P.",
    title = "{Exotic baryons in hot neutron stars}",
    eprint = "2302.04364",
    archivePrefix = "arXiv",
    primaryClass = "nucl-th",
    doi = "10.1093/mnras/stad1198",
    journal = "Mon. Not. Roy. Astron. Soc.",
    volume = "522",
    number = "3",
    pages = "3263--3270",
    year = "2023"
}

@article{Prakash:1996xs,
    author = "Prakash, Madappa and Bombaci, Ignazio and Prakash, Manju and Ellis, Paul J. and Lattimer, James M. and Knorren, Roland",
    title = "{Composition and structure of protoneutron stars}",
    eprint = "nucl-th/9603042",
    archivePrefix = "arXiv",
    reportNumber = "SUNY-NTG-96-11, NUC-MINN-93-23-T",
    doi = "10.1016/S0370-1573(96)00023-3",
    journal = "Phys. Rept.",
    volume = "280",
    pages = "1--77",
    year = "1997"
}

@article{Pons:1998mm,
    author = "Pons, J. A. and Reddy, S. and Prakash, M. and Lattimer, J. M. and Miralles, J. A.",
    title = "{Evolution of protoneutron stars}",
    eprint = "astro-ph/9807040",
    archivePrefix = "arXiv",
    reportNumber = "SUNY-NTG-98-30",
    doi = "10.1086/306889",
    journal = "Astrophys. J.",
    volume = "513",
    pages = "780",
    year = "1999"
}

@article{Kumari:2021tik,
    author = "Kumari, Manisha and Kumar, Arvind",
    title = "{Properties of strange quark matter and strange quark stars}",
    doi = "10.1140/epjc/s10052-021-09576-w",
    journal = "Eur. Phys. J. C",
    volume = "81",
    number = "9",
    pages = "791",
    year = "2021"
}

@article{Backes:2020fyw,
    author = "Backes, B. C. and Hafemann, E. and Marzola, I. and Menezes, D. P.",
    title = "{Density dependent quark mass model revisited: Thermodynamic consistency, stability windows and stellar properties}",
    eprint = "2007.04494",
    archivePrefix = "arXiv",
    primaryClass = "hep-ph",
    doi = "10.1088/1361-6471/abc6e9",
    journal = "J. Phys. G",
    volume = "48",
    number = "5",
    pages = "055104",
    year = "2021"
}

@article{daSilva:2023okq,
    author = "da Silva, Franciele M. and Issifu, Adamu and Lopes, Luiz L. and Santos, Luis C. N. and Menezes, D{\'e}bora P.",
    title = "{Bayesian study of quark models in view of recent astrophysical constraints}",
    eprint = "2309.16865",
    archivePrefix = "arXiv",
    primaryClass = "nucl-th",
    doi = "10.1103/PhysRevD.109.043054",
    journal = "Phys. Rev. D",
    volume = "109",
    number = "4",
    pages = "043054",
    year = "2024"
}

@article{Issifu_2025,
doi = {10.1088/1361-6382/ade192},
url = {https://doi.org/10.1088/1361-6382/ade192},
year = {2025},
month = {jun},
publisher = {IOP Publishing},
volume = {42},
number = {12},
pages = {125004},
author = {Issifu, Adamu and da Silva, Franciele M and Santos, Luis C N and Menezes, Débora P and Frederico, Tobias},
title = {Strongly interacting quark matter in massive quark stars},
journal = {Classical and Quantum Gravity}
}

@article{Xia:2014zaa,
    author = "Xia, C. J. and Peng, G. X. and Chen, S. W. and Lu, Z. Y. and Xu, J. F.",
    title = "{Thermodynamic consistency, quark mass scaling, and properties of strange matter}",
    eprint = "1405.3037",
    archivePrefix = "arXiv",
    primaryClass = "hep-ph",
    doi = "10.1103/PhysRevD.89.105027",
    journal = "Phys. Rev. D",
    volume = "89",
    number = "10",
    pages = "105027",
    year = "2014"
}

@article{Alford:2004zr,
    author = "Alford, Mark and Jotwani, Pooja and Kouvaris, Chris and Kundu, Joydip and Rajagopal, Krishna",
    title = "{A Hot water bottle for aging neutron stars}",
    eprint = "astro-ph/0411560",
    archivePrefix = "arXiv",
    reportNumber = "MIT-CTP-3558, UMPP-05-022",
    doi = "10.1103/PhysRevD.71.114011",
    journal = "Phys. Rev. D",
    volume = "71",
    pages = "114011",
    year = "2005"
}

@article{Weber:2004kj,
    author = "Weber, Fridolin",
    title = "{Strange quark matter and compact stars}",
    eprint = "astro-ph/0407155",
    archivePrefix = "arXiv",
    doi = "10.1016/j.ppnp.2004.07.001",
    journal = "Prog. Part. Nucl. Phys.",
    volume = "54",
    pages = "193--288",
    year = "2005"
}

@article{Drago:2015dea,
    author = "Drago, Alessandro and Pagliara, Giuseppe",
    title = "{The scenario of two families of compact stars}: {2. Transition from hadronic to quark matter and explosive phenomena}",
    eprint = "1509.02134",
    archivePrefix = "arXiv",
    primaryClass = "astro-ph.SR",
    doi = "10.1140/epja/i2016-16041-2",
    journal = "Eur. Phys. J. A",
    volume = "52",
    number = "2",
    pages = "41",
    year = "2016"
}

@article{Paschalidis:2016vmz,
    author = "Paschalidis, Vasileios and Stergioulas, Nikolaos",
    title = "{Rotating Stars in Relativity}",
    eprint = "1612.03050",
    archivePrefix = "arXiv",
    primaryClass = "astro-ph.HE",
    doi = "10.1007/s41114-017-0008-x",
    journal = "Living Rev. Rel.",
    volume = "20",
    number = "1",
    pages = "7",
    year = "2017"
}

@ARTICLE{2022NatAs...6.1444D,
       author = {{Doroshenko}, Victor and {Suleimanov}, Valery and {P{\"u}hlhofer}, Gerd and {Santangelo}, Andrea},
        title = "{A strangely light neutron star within a supernova remnant}",
      journal = {Nature Astronomy},
         year = 2022,
        month = dec,
       volume = {6},
        pages = {1444-1451},
          doi = {10.1038/s41550-022-01800-1},
       adsurl = {https://ui.adsabs.harvard.edu/abs/2022NatAs...6.1444D}
}

@article{Drake:2002bj,
    author = "Drake, Jeremy J. and others",
    title = "{Is RXJ1856.5-3754 a quark star?}",
    eprint = "astro-ph/0204159",
    archivePrefix = "arXiv",
    doi = "10.1086/340368",
    journal = "Astrophys. J.",
    volume = "572",
    pages = "996--1001",
    year = "2002"
}

@Article{ZhouPhRvD2018,
  author        = {Zhou, Enping and Tsokaros, Antonios and Rezzolla, Luciano and Xu, Renxin and Ury{\={u}}, K{\={o}}ji},
  journal       = {\prd},
  title         = {{Uniformly rotating, axisymmetric, and triaxial quark stars in general relativity}},
  year          = {2018},
  month         = jan,
  number        = {2},
  pages         = {023013},
  volume        = {97},
  adsurl        = {https://ui.adsabs.harvard.edu/abs/2018PhRvD..97b3013Z},
  archiveprefix = {arXiv},
  doi           = {10.1103/PhysRevD.97.023013},
  eid           = {023013},
  primaryclass  = {astro-ph.HE},
}

@ARTICLE{1994ApJ...424..823C,
       author = {{Cook}, Gregory B. and {Shapiro}, Stuart L. and {Teukolsky}, Saul A.},
        title = "{Rapidly Rotating Neutron Stars in General Relativity: Realistic Equations of State}",
      journal = {\apj},
         year = 1994,
        month = apr,
       volume = {424},
        pages = {823},
          doi = {10.1086/173934},
       adsurl = {https://ui.adsabs.harvard.edu/abs/1994ApJ...424..823C}
}

@Article{komatsu1989MNRASa,
  author        = {Komatsu, H. and Eriguchi, Y. and Hachisu, I.},
  title         = {Rapidly rotating general relativistic stars. I -- Numerical method and its application to uniformly rotating polytropes},
  journal       = {Mon. Not. R. Astron. Soc.},
  year          = {1989},
  month         = mar,
  volume        = {237},
  pages         = {355--379},
  doi           = {10.1093/mnras/237.2.355},
  adsurl        = {https://ui.adsabs.harvard.edu/abs/1989MNRAS.237..355K}
}

@article{Burrows:1986me,
    author = "Burrows, Adam and Lattimer, James M.",
    title = "{The birth of neutron stars}",
    doi = "10.1086/164405",
    journal = "Astrophys. J.",
    volume = "307",
    pages = "178--196",
    year = "1986"
}

@BOOK{1996csnp.book.....G,
       author = {{Glendenning}, Norman},
        title = "{Compact Stars. Nuclear Physics, Particle Physics and General Relativity.}",
         year = 1996,
       adsurl = {https://ui.adsabs.harvard.edu/abs/1996csnp.book.....G}
}

@article{Stergioulas:2003yp,
    author = "Stergioulas, Nikolaos",
    title = "{Rotating Stars in Relativity}",
    eprint = "gr-qc/0302034",
    archivePrefix = "arXiv",
    reportNumber = "AUTH-LR2003",
    doi = "10.12942/lrr-2003-3",
    journal = "Living Rev. Rel.",
    volume = "6",
    pages = "3",
    year = "2003"
}

@article{Choudhury:2024xbk,
    author = "Choudhury, Devarshi and others",
    title = "{A NICER View of the Nearest and Brightest Millisecond Pulsar: PSR J0437\textendash{}4715}",
    doi = "10.3847/2041-8213/ad5a6f",
    journal = "Astrophys. J. Lett.",
    volume = "971",
    number = "1",
    pages = "L20",
    year = "2024"
}

@article{Gondek-Rosinska:2001kuc,
    author = "Gondek-Rosinska, Dorota and Stergioulas, Nikolaos and Bulik, Tomasz and Kluzniak, Wlodek and Gourgoulhon, Eric",
    title = "{Lower limits on the maximum orbital frequency around rotating strange stars}",
    eprint = "astro-ph/0110209",
    archivePrefix = "arXiv",
    doi = "10.1051/0004-6361:20011328",
    journal = "Astron. Astrophys.",
    volume = "380",
    pages = "190--197",
    year = "2001"
}

@article{miller2021,
    doi = {10.3847/2041-8213/ac089b},
    url = {https://dx.doi.org/10.3847/2041-8213/ac089b},
year = {2021},
month = {sep},
publisher = {The American Astronomical Society},
volume = {918},
number = {2},
pages = {L28},
author = {M. C. Miller and others},
title = {The Radius of PSR J0740+6620 from NICER and XMM-Newton Data},
journal = {The Astrophysical Journal Letters}
}

@article{PhysRevD.30.2379,
  author = {Farhi, Edward and Jaffe, R. L.},
  journal = {Phys. Rev. D},
  volume = {\textbf{30}},
  issue = {11},
  pages = {2379},
  numpages = {0},
  year = {1984},
  publisher = {American Physical Society},
  doi = {10.1103/PhysRevD.30.2379},
  url = {https://link.aps.org/doi/10.1103/PhysRevD.30.2379}
}

@InProceedings{SzkudlarekASPC2012, 
  author    = {Szkudlarek, M. and Gondek-Rosi{\'n}ska, D. and Villain, L. and Ansorg, M.},
  booktitle = {Electromagnetic Radiation from Pulsars and Magnetars},
  title     = {{The Maximum Mass of Rotating Strange Stars}},
  year      = {2012},
  editor    = {{Lewandowski}, W. and {Maron}, O. and {Kijak}, J.},
  month     = dec,
  pages     = {231},
  series    = {Astronomical Society of the Pacific Conference Series},
  volume    = {466},
  adsurl    = {https://ui.adsabs.harvard.edu/abs/2012ASPC..466..231S},
}

@article{Ouyed_2002,
   title={Quark-Nova},
   volume={390},
   ISSN={1432-0746},
   url={http://dx.doi.org/10.1051/0004-6361:20020982},
   DOI={10.1051/0004-6361:20020982},
   number={3},
   journal={Astronomy \& Astrophysics},
   publisher={EDP Sciences},
   author={Ouyed, R. and Dey, J. and Dey, M.},
   year={2002},
   month={Aug},
   pages={L39–L42}
}

@ARTICLE{Stergioulas95,
       author = {{Stergioulas}, Nikolaos and {Friedman}, John L.},
        title = "{Comparing Models of Rapidly Rotating Relativistic Stars Constructed by Two Numerical Methods}",
      journal = {Astrophys. J.},
         year = 1995,
        month = may,
       volume = {444},
        pages = {306},
          doi = {10.1086/175605},
archivePrefix = {arXiv},
       eprint = {astro-ph/9411032},
 primaryClass = {astro-ph},
       adsurl = {https://ui.adsabs.harvard.edu/abs/1995ApJ...444..306S}
}

@article{DiClemente:2022wqp,
    author = "Di Clemente, Francesco and Drago, Alessandro and Pagliara, Giuseppe",
    title = "{Is the Compact Object Associated with HESS J1731-347 a Strange Quark Star? A Possible Astrophysical Scenario for Its Formation}",
    eprint = "2211.07485",
    archivePrefix = "arXiv",
    primaryClass = "astro-ph.HE",
    doi = "10.3847/1538-4357/ad445b",
    journal = "Astrophys. J.",
    volume = "967",
    number = "2",
    pages = "159",
    year = "2024"
}

@article{Li:1999mk,
    author = "Li, Xiang-Dong and Ray, Subharthi and Dey, Jishnu and Dey, Mira and Bombaci, Ignazio",
    title = "{On the Nature of the compact star in 4u 1728-34}",
    eprint = "astro-ph/9908274",
    archivePrefix = "arXiv",
    doi = "10.1086/312394",
    journal = "Astrophys. J. Lett.",
    volume = "527",
    pages = "L51--L54",
    year = "1999"
}

@article{Miller:2019cac,
    author = "Miller, M.C. and others",
    title = "{PSR J0030+0451 Mass and Radius from $NICER$ Data and Implications for the Properties of Neutron Star Matter}",
    eprint = "1912.05705",
    archivePrefix = "arXiv",
    primaryClass = "astro-ph.HE",
    doi = "10.3847/2041-8213/ab50c5",
    journal = "Astrophys. J. Lett.",
    volume = "887",
    number = "1",
    pages = "L24",
    year = "2019"
}

@Article{PrakashPhLB1990e,
  author   = {Prakash, M. and Baron, E. and Prakash, M.},
  journal  = {Phys. Lett. B},
  title    = {{Erratum: Rotation of stars containing strange quark matter [Phys. Lett. B 243 (1990) 175]}},
  year     = {1990},
  month    = sep,
  number   = {4},
  pages    = {632--632},
  volume   = {247},
  adsurl   = {https://ui.adsabs.harvard.edu/abs/1990PhLB..247..632P},
  doi      = {10.1016/0370-2693(90)91913-V},
  fjournal = {Physics Letters B},
}

@Article{ZdunikPhRvD1990,
  author  = {Zdunik, J.~L. and Haensel, P.},
  journal = {\prd},
  title   = {{Maximum rotation frequency of strange stars}},
  year    = {1990},
  month   = jul,
  number  = {2},
  pages   = {710--711},
  volume  = {42},
  adsurl  = {https://ui.adsabs.harvard.edu/abs/1990PhRvD..42..710Z},
  doi     = {10.1103/PhysRevD.42.710},
}

@article{Deur:2016tte,
    author = "Deur, Alexandre and Brodsky, Stanley J. and de Teramond, Guy F.",
    title = "{The QCD Running Coupling}",
    eprint = "1604.08082",
    archivePrefix = "arXiv",
    primaryClass = "hep-ph",
    reportNumber = "JLAB-PHY-16-2199, SLAC-PUB-16448, DOE/OR/23177-3645, DOE-OR-23177-3645",
    doi = "10.1016/j.ppnp.2016.04.003",
    journal = "Nucl. Phys.",
    volume = "90",
    pages = "1",
    year = "2016"
}

@article{ParticleDataGroup:2022pth,
    author = "Workman, R. L. and others",
    collaboration = "Particle Data Group",
    title = "{Review of Particle Physics}",
    doi = "10.1093/ptep/ptac097",
    journal = "PTEP",
    volume = "2022",
    pages = "083C01",
    year = "2022"
}

@article{Drago:2013fsa,
    author = "Drago, Alessandro and Lavagno, Andrea and Pagliara, Giuseppe",
    title = "{Can very compact and very massive neutron stars both exist?}",
    eprint = "1309.7263",
    archivePrefix = "arXiv",
    primaryClass = "nucl-th",
    doi = "10.1103/PhysRevD.89.043014",
    journal = "Phys. Rev. D",
    volume = "89",
    number = "4",
    pages = "043014",
    year = "2014"
}

@article{Lugones:2002zd,
    author = "Lugones, G. and Horvath, J. E.",
    title = "{High-density QCD pairing in compact star structure}",
    eprint = "astro-ph/0211638",
    archivePrefix = "arXiv",
    doi = "10.1051/0004-6361:20030374",
    journal = "Astron. Astrophys.",
    volume = "403",
    pages = "173--178",
    year = "2003"
}

@Article{universe8060322,
AUTHOR = {Ouyed, Rachid},
TITLE = {The Macro-Physics of the Quark-Nova: Astrophysical Implications},
JOURNAL = {Universe},
VOLUME = {8},
YEAR = {2022},
NUMBER = {6},
ARTICLE-NUMBER = {322},
URL = {https://www.mdpi.com/2218-1997/8/6/322},
ISSN = {2218-1997},
DOI = {10.3390/universe8060322}
}

@article{Ouyed_2020,
doi = {10.1088/1674-4527/20/2/27},
url = {https://doi.org/10.1088/1674-4527/20/2/27},
year = {2020},
month = {mar},
publisher = {National Astronomical Observatories, CAS and IOP Publishing Ltd.},
volume = {20},
number = {2},
pages = {027},
author = {Ouyed, Rachid and Leahy, Denis and Koning, Nico},
title = {A quark nova in the wake of a core-collapse supernova: a unifying model for long duration gamma-ray bursts and fast radio bursts},
journal = {Research in Astronomy and Astrophysics}
}

@article{Ouyed:2001cg,
    author = "Ouyed, Rachid and Sannino, Francesco",
    title = "{Quark stars as inner engines for Gamma ray bursts?}",
    eprint = "astro-ph/0103022",
    archivePrefix = "arXiv",
    doi = "10.1051/0004-6361:20020409",
   journal = {Astronomy \& Astrophysics},
    volume = "387",
    pages = "725",
    year = "2002"
}

@article{Ouyed:2001ts,
    author = "Ouyed, Rachid and Dey, Jishnu and Dey, Mira",
    title = "{Quark - nova as gamma-ray burst precursor}",
    eprint = "astro-ph/0105109",
    archivePrefix = "arXiv",
    doi = "10.1051/0004-6361:20020982",
    journal = {Astronomy \& Astrophysics},
    volume = "390",
    pages = "L39",
    year = "2002"
}

@Article{BhattacharyyaMNRAS2016,
   author = "Bhattacharyya, Sudip and Bombaci, Ignazio and Logoteta, Domenico and Thampan, Arun V.",
    title = "{Fast spinning strange stars: possible ways to constrain interacting quark matter parameters}",
    eprint = "1601.06120",
    archivePrefix = "arXiv",
    primaryClass = "astro-ph.HE",
    doi = "10.1093/mnras/stw206",
    journal = "Mon. Not. Roy. Astron. Soc.",
    volume = "457",
    number = "3",
    pages = "3101--3114",
    year = "2016"
}

@Article{Gondek-RosinskaA&A2000,
  author        = {Gondek-Rosi{\'n}ska, D. and Bulik, T. and Zdunik, J. L. and Gourgoulhon, E. and Ray, S. and Dey, J. and Dey, M.},
  title         = {Rapidly rotating compact strange stars},
  journal       = {Astronomy and Astrophysics},
  year          = {2000},
  volume        = {363},
  pages         = {1005--1012},
  month         = nov,
  doi           = {10.48550/arXiv.astro-ph/0007004},
  archivePrefix = {arXiv},
  eprint        = {astro-ph/0007004},
  primaryClass  = {astro-ph},
  adsurl        = {https://ui.adsabs.harvard.edu/abs/2000A&A...363.1005G}
}

@article{Konstantinou_2026,
doi = {10.3847/1538-4357/ae2f62},
url = {https://doi.org/10.3847/1538-4357/ae2f62},
year = {2026},
month = {jan},
publisher = {The American Astronomical Society},
volume = {997},
number = {1},
pages = {55},
author = {Konstantinou, Andreas},
title = {The Effect of a Self-bound Equation of State on the Structure of Rotating Compact Stars},
journal = {The Astrophysical Journal}
}

@article{eXTP:2018anb,
    author = "Zhang, Shuang-Nan and others",
    collaboration = "eXTP",
    title = "{The enhanced X-ray Timing and Polarimetry mission{\textemdash}eXTP}",
    eprint = "1812.04020",
    archivePrefix = "arXiv",
    primaryClass = "astro-ph.IM",
    doi = "10.1007/s11433-018-9309-2",
    journal = "Sci. China Phys. Mech. Astron.",
    volume = "62",
    number = "2",
    pages = "29502",
    year = "2019"
}

@article{Mashhoon:2006fj,
    author = "Mashhoon, Bahram and Singh, Dinesh",
    title = "{Dynamics of Extended Spinning Masses in a Gravitational Field}",
    eprint = "astro-ph/0608278",
    archivePrefix = "arXiv",
    doi = "10.1103/PhysRevD.74.124006",
    journal = "Phys. Rev. D",
    volume = "74",
    pages = "124006",
    year = "2006"
}

@article{watts2019dense,
  title={Dense matter with eXTP},
  author={Watts, Anna L and Yu, WenFei and Poutanen, Juri and Zhang, Shu and Bhattacharyya, Sudip and Bogdanov, Slavko and Ji, Long and Patruno, Alessandro and Riley, Thomas E and Bakala, Pavel and others},
  journal={Science China Physics, Mechanics \& Astronomy},
  volume={62},
  number={2},
  pages={29503},
  year={2019},
  publisher={Springer}
}

@article{Konstantinou:2022vkr,
    title={Universal Relations for the Increase in the Mass and Radius of a Rotating Neutron Star},
   volume={934},
   ISSN={1538-4357},
   url={http://dx.doi.org/10.3847/1538-4357/ac7b86},
   DOI={10.3847/1538-4357/ac7b86},
   number={2},
   journal={The Astrophysical Journal},
   publisher={American Astronomical Society},
   author={Konstantinou, Andreas and Morsink, Sharon M.},
   year={2022},
   month=aug, pages={139} }

@article{riley2021,
title={A NICER View of the Massive Pulsar PSR J0740+6620 Informed by Radio Timing and XMM-Newton Spectroscopy},
   volume={918},
   ISSN={2041-8213},
   url={http://dx.doi.org/10.3847/2041-8213/ac0a81},
   number={2},
   journal={Astrophys. J. Lett.},
   publisher={American Astronomical Society},
   author = {Riley {\sl et al.}, T. E. },
   year = {2021}
}

@article{Lattimer:2000nx,
    author = "Lattimer, J. M. and Prakash, M.",
    title = "{Neutron star structure and the equation of state}",
    eprint = "astro-ph/0002232",
    archivePrefix = "arXiv",
    doi = "10.1086/319702",
    journal = "Astrophys. J.",
    volume = "550",
    pages = "426",
    year = "2001"
}

@article{Witten:1984rs,
    author = "Witten, Edward",
    title = "{Cosmic Separation of Phases}",
    reportNumber = "PRINT-84-0400 (IAS,PRINCETON)",
    doi = "10.1103/PhysRevD.30.272",
    journal = "Phys. Rev. D",
    volume = "30",
    pages = "272--285",
    year = "1984"
}

@ARTICLE{1986ApJ...310..261A,
       author = {{Alcock}, Charles and {Farhi}, Edward and {Olinto}, Angela},
        title = "{Strange Stars}",
      journal = {\apj},
         year = 1986,
        month = nov,
       volume = {310},
        pages = {261},
          doi = {10.1086/164679},
       adsurl = {https://ui.adsabs.harvard.edu/abs/1986ApJ...310..261A}
}

@article{Bodmer:1971we,
    author = "Bodmer, A. R.",
    title = "{Collapsed nuclei}",
    doi = "10.1103/PhysRevD.4.1601",
    journal = "Phys. Rev. D",
    volume = "4",
    pages = "1601--1606",
    year = "1971"
}

@article{Lattimer:2004pg,
    author = "Lattimer, J. M. and Prakash, M.",
    title = "{The physics of neutron stars}",
    eprint = "astro-ph/0405262",
    archivePrefix = "arXiv",
    doi = "10.1126/science.1090720",
    journal = "Science",
    volume = "304",
    pages = "536--542",
    year = "2004"
}

@article{Oertel:2016bki,
    author = {Oertel, M. and Hempel, M. and Kl\"ahn, T. and Typel, S.},
    title = "{Equations of state for supernovae and compact stars}",
    eprint = "1610.03361",
    archivePrefix = "arXiv",
    primaryClass = "astro-ph.HE",
    doi = "10.1103/RevModPhys.89.015007",
    journal = "Rev. Mod. Phys.",
    volume = "89",
    number = "1",
    pages = "015007",
    year = "2017"
}

@nline{Issifu:2025hqg,
    author = "Issifu, Adamu and Konstantinou, Andreas and da Silva, Franciele M. and Frederico, Tobias",
    title = "{Rotational effects in quark stars: comparing different models}",
    eprint = "2511.20477",
    archivePrefix = "arXiv",
    primaryClass = "astro-ph.HE",
    month = "11",
    year = "2025"
}

@article{Will:2018ont,
    author = "Will, Clifford M.",
    title = "{Testing general relativity with compact-body orbits: a modified Einstein{\textendash}Infeld{\textendash}Hoffmann framework}",
    eprint = "1801.08999",
    archivePrefix = "arXiv",
    primaryClass = "gr-qc",
    doi = "10.1088/1361-6382/aab1c6",
    journal = "Class. Quant. Grav.",
    volume = "35",
    number = "8",
    pages = "085001",
    year = "2018"
}

@article{Yagi:2016bkt,
    author = "Yagi, Kent and Yunes, Nicol\'as",
    title = "{Approximate Universal Relations for Neutron Stars and Quark Stars}",
    eprint = "1608.02582",
    archivePrefix = "arXiv",
    primaryClass = "gr-qc",
    doi = "10.1016/j.physrep.2017.03.002",
    journal = "Phys. Rept.",
    volume = "681",
    pages = "1--72",
    year = "2017"
}

@article{Riley2019,
author = {Riley {\sl et al.}, T.E. },
journal = {Astrophys. J. Lett.},
number = {1},
pages = {L21},
title = {{A $NICER$ View of PSR J0030+0451: Millisecond Pulsar Parameter Estimation}},
volume = {887},
year = {2019}
}

@article{Silva:2020acr,
    author = "Silva, Hector O. and Holgado, A. Miguel and C{\'a}rdenas-Avenda{\~n}o, Alejandro and Yunes, Nicol{\'a}s",
    title = "{Astrophysical and theoretical physics implications from multimessenger neutron star observations}",
    eprint = "2004.01253",
    archivePrefix = "arXiv",
    primaryClass = "gr-qc",
    doi = "10.1103/PhysRevLett.126.181101",
    journal = "Phys. Rev. Lett.",
    volume = "126",
    number = "18",
    pages = "181101",
    year = "2021"
}

@article{Peng:2000ff,
    author = "Peng, G. X. and Chiang, H. C. and Ning, P. Z.",
    title = "{Thermodynamics, strange quark matter, and strange stars}",
    eprint = "hep-ph/0003027",
    archivePrefix = "arXiv",
    doi = "10.1103/PhysRevC.62.025801",
    journal = "Phys. Rev. C",
    volume = "62",
    pages = "025801",
    year = "2000"
}

@article{Li_2022,
doi = {10.1088/1361-6382/ac45d9},
url = {https://dx.doi.org/10.1088/1361-6382/ac45d9},
year = {2022},
month = {jan},
publisher = {IOP Publishing},
volume = {39},
number = {3},
pages = {035014},
author = {Li, Yuxi and Wang, Jue and Wu, Zehan and Wen, Dehua},
title = {Inferring the gravitational binding energy and moment of inertia of PSR J0030 + 0451 and PSR J0740 + 6620 from new universal relations},
journal = {Classical and Quantum Gravity}
}

@article{Kumar:2019xgp,
    author = "Kumar, Bharat and Landry, Philippe",
    title = "{Inferring neutron star properties from GW170817 with universal relations}",
    eprint = "1902.04557",
    archivePrefix = "arXiv",
    primaryClass = "gr-qc",
    doi = "10.1103/PhysRevD.99.123026",
    journal = "Phys. Rev. D",
    volume = "99",
    number = "12",
    pages = "123026",
    year = "2019"
}

@article{PhysRevD.105.063023,
  title = {Exploring the universal relations with the correlation analysis of neutron star properties},
  author = {Yang, Shen and Wen, Dehua and Wang, Jue and Zhang, Jing},
  journal = {Phys. Rev. D},
  volume = {105},
  issue = {6},
  pages = {063023},
  numpages = {9},
  year = {2022},
  month = {Mar},
  publisher = {American Physical Society},
  doi = {10.1103/PhysRevD.105.063023},
  url = {https://link.aps.org/doi/10.1103/PhysRevD.105.063023}
}

@article{Bejger:2005jy,
    author = "Bejger, Michal and Bulik, T. and Haensel, P.",
    title = "{Moments of inertia of the binary pulsars J0737-3039A,B and the dense matter EOS}",
    eprint = "astro-ph/0508105",
    archivePrefix = "arXiv",
    doi = "10.1111/j.1365-2966.2005.09575.x",
    journal = "Mon. Not. Roy. Astron. Soc.",
    volume = "364",
    pages = "635",
    year = "2005"
}

\end{document}